\documentclass[%
 reprint,
superscriptaddress,
showpacs,preprintnumbers,
nofootinbib,
 amsmath,amssymb,
 aps,
prd,
]{revtex4-1}
\usepackage{graphicx}
\usepackage{dcolumn}
\usepackage{bm}
\usepackage{lipsum}
\usepackage{color}
\usepackage{grffile}
\usepackage{amsmath}
\usepackage{mathtools}
\usepackage{amssymb}
\usepackage{xfrac}
\usepackage{fancyhdr}
\usepackage{enumerate}
\usepackage{indentfirst}
\usepackage[hang,flushmargin]{footmisc} 
\usepackage{bibentry}
\usepackage{xspace}

\newcommand{\numu}{\ensuremath{\nu_{\mu}}\xspace}
\newcommand{\nue}{\ensuremath{\nu_{e}}\xspace}
\newcommand{\numubar}{$\ensuremath{\bar{\nu}_{\mu}}\xspace$}
\newcommand{\nuebar}{$\ensuremath{\bar{\nu}_{e}}\xspace$}

\newcommand{\chisq}{$\ensuremath{\chi^2}\xspace$}

\begin{document}

\preprint{IPPP/17/11}

\title{Prospects of Light Sterile Neutrino Oscillation and CP Violation Searches at the Fermilab Short Baseline Neutrino Facility}

\author{D.~Cianci}
\affiliation{
Department of Physics, Columbia University, New York, NY 10027, USA
}
\affiliation{
School of Physics and Astronomy,  The University of Manchester, Manchester, M13 9PL, UK
}
\author{A.~Furmanski}
\affiliation{
School of Physics and Astronomy,  The University of Manchester, Manchester, M13 9PL, UK
}
\author{G.~Karagiorgi}
\affiliation{
Department of Physics, Columbia University, New York, NY 10027, USA
}
\affiliation{
School of Physics and Astronomy,  The University of Manchester, Manchester, M13 9PL, UK
} 
\author{M.~Ross-Lonergan}
\affiliation{
Institute of Particle Physics Phenomenology, Durham University, Durham, DH1 3LE, UK
}

\date{\today}

\begin{abstract}
    We investigate the ability of the Short Baseline Neutrino (SBN) experimental program at Fermilab to test the globally-allowed (3+$N$) sterile neutrino oscillation parameter space. We explicitly consider the globally-allowed parameter space for the (3+1), (3+2), and (3+3) sterile neutrino oscillation scenarios. We find that SBN can probe with $5\sigma$ sensitivity more than 85\%, 95\% and 55\% of the parameter space currently allowed at 99\% confidence level for the (3+1), (3+2) and (3+3) scenarios, respectively. In the case of the (3+2) and (3+3) scenarios, CP-violating phases appear in the oscillation probability terms, leading to observable differences in the appearance probabilities of neutrinos and antineutrinos. We explore SBN's sensitivity to those phases for the (3+2) scenario through the currently planned neutrino beam running, and investigate potential improvements through additional antineutrino beam running. We show that, if antineutrino exposure is considered, for maximal values of the (3+2) CP-violating phase $\phi_{54}$, SBN could be the first experiment to directly observe $\sim 2\sigma$ hints of CP violation associated with an extended lepton sector.
\end{abstract}

\maketitle


\section{\label{sec:intro}Introduction}

During the past few decades, concurrently with the experimental confirmation of three-neutrino oscillations, several additional oscillation-like anomalous experimental signatures have surfaced, which may require new physics to interpret. One possible such new physics interpretation is that of additional, light sterile neutrinos \cite{Abazajian:2012ys}. Those are new neutrino states which are assumed to have no weak interactions and are associated with light neutrino masses of order $0.1-10$~eV. The mass states are thought to have small weak flavor content (electron, muon, and potentially tau), leading to small-amplitude neutrino oscillations at relatively small $L/E\sim1$~m/MeV. The constraint of small weak flavor content (in particular electron and muon flavor) is imposed by unitarity of the overall neutrino mixing matrix, together with existing experimental bounds on the elements of the neutrino mixing matrix (see, e.g.,~\cite{Parke:2015goa,Collin:2016aqd}). The $L/E$ over which such oscillations manifest is what dictates the associated mass splittings of $0.1^2-10^2$~eV$^2$. This signature is often referred to as short-baseline oscillations.

These anomalous short-baseline oscillation observations are contributed primarily by the LSND \cite{ds:lsnd} and MiniBooNE \cite{Aguilar-Arevalo:2013pmq} experiments. Both experiments have searched for $\nu_e$ appearance in a $\nu_{\mu}$-dominated beam, at a similar $L/E$, albeit each at a different $L$, and with a different neutrino beam energy, $E$. A third observation consistent with short-baseline oscillations has been provided in the $\nu_e$ disappearance channel from calibration measurements employing intense radioactive sources of high $\nu_e$ flux in radiochemical experiments, during the mid 1980's \cite{ds:gal,ds:gal2}. A fourth hint had been provided by past reactor-based short-baseline oscillation searches; specifically, recent reactor data re-analyses using updated reactor flux predictions showed evidence of a deficit in the reactor electron antineutrino event rates measured collectively by several experiments at $L/E$ values ranging between 2-20~m/MeV. This has been referred to as the ``reactor anomaly''~\cite{Mention:2011rk}. However, recent realizations that large and unaccounted-for systematic uncertainties are associated with reactor neutrino flux predictions (see, e.g.~\cite{Dwyer:2015pgm,Hayes:2013wra,Hayes:2015ctl}) dictate that the reactor anomaly cannot yet be interpreted decisively as a light sterile neutrino oscillation signature; such interpretations should await either improved reactor antineutrino flux modelling or dedicated searches for light sterile neutrino oscillations at reactor short baselines which are sensitive to distortions in reconstructed event spectra that are $L/E$ dependent. Such searches are now under way with a number of experiments \cite{Alekseev:2016llm,Ko:2016owz,Serebrov:2012sq,Lane:2015alq,Ashenfelter:2013oaa,Helaine:2016bmc,Ryder:2015sma}.

Interpreting the above $\nu_{\mu}\rightarrow\nu_e$ appearance and $\nu_e$ disappearance observations as sterile neutrino oscillations would imply large $\nu_{\mu}$ disappearance observable at short baselines. Such signature has not yet been observed; on the contrary, multiple experiments have imposed stringent bounds on sterile neutrino mixing parameters involved in the $\nu_\mu$ disappearance channel, bringing the viability of sterile neutrino models into question \cite{Giunti:2015jvi}. The most recent $\nu_{\mu}$ disappearance data sets include IceCube~\cite{TheIceCube:2016oqi} and MINOS+~\cite{MINOS:2016viw}. The most up to date global fits and results, incorporating IceCube constraints, are presented in Ref.~\cite{Collin:2016rao}. Despite the strong disappearance constraints, the MiniBooNE, LSND, and arguably the calibration source experimental results still stand as anomalous observations that require further investigation to resolve.

To definitively address these collective anomalies, the Short Baseline Neutrino (SBN) experimental program~\cite{Antonello:2015lea} was successfully proposed and is now under construction in the Booster Neutrino Beamline (BNB) at Fermilab. The BNB provides a high intensity, sign-selected, primarily ($>$99\%) muon neutrino (and muon antineutrino) flux~\cite{ds:mbnu}. Three liquid argon time projection chamber (LArTPC) detectors, comprising the already operating MicroBooNE detector, the SBND detector which is under construction, and the ICARUS detector which is under refurbishment, sample the $\nu_e$ and $\nu_{\mu}$ flux content at three distinct baselines. This allows SBN to perform electron neutrino appearance and muon neutrino disappearance searches with highly competitive sensitivity coverage, as presented in the SBN proposal \cite{Antonello:2015lea}. Note, however, that the discovery potential of SBN has only been considered for the simplest sterile neutrino oscillation scenario, where only a single additional, mostly sterile neutrino mass eigenstate is assumed; this scenario is referred to as a (3+1) scenario.

In this paper, we perform an independent phenomenological study where we expand beyond the (3+1) scenario and, for the first time, evaluate SBN's sensitivity to sterile neutrino oscillation models with two and three additional sterile neutrinos, referred to as (3+2) and (3+3), respectively. Furthermore, for the (3+1) scenario, we re-evaluate SBN's sensitivity to electron neutrino appearance without the explicit assumption of negligible disappearance of intrinsic $\nu_e$ backgrounds, extending beyond what has been followed by the SBN collaboration in \cite{Antonello:2015lea}. Finally, for the (3+2) scenario, we explore SBN's sensitivity to additional CP violation that is potentially observable under this oscillation assumption. Although currently SBN is only approved to run in neutrino mode, it is interesting to consider what potential antineutrino mode running could add in terms of sensitivity to CP violation phases. We explore this question more explicitly for added antineutrino beam running at SBN, beyond the presently planned neutrino running.

The large (3+$N$) parameter space dimensionality for $N=2,3$ makes it particularly challenging to provide simple and meaningful quantitative statements on SBN's sensitivity reach with respect to these models. To deal with this issue, we have devised a new sensitivity metric that exploits existing experimental constraints to sterile neutrino oscillation scenarios to effectively reduce the parameter space over which SBN's sensitivity reach must be quantified. The constraints are provided in the form of global fits to a representative sample of short-baseline oscillation data sets (both signal and null results), and are used to define a hypervolume of allowed parameter space under each (3+$N$) hypothesis over which SBN's sensitivity is evaluated.

The paper is organized as follows: In Sec.~\ref{sec:formalism}, we introduce the sterile neutrino oscillation formalism followed in this work. In Sec.~\ref{sec:globalfit} we give the prescription used to fit global sterile neutrino oscillation data to reduce the parameter space over which SBN's sensitivity is evaluated; we also summarize the results of fits performed under each oscillation hypothesis in Secs.~\ref{sec:threeone}-\ref{sec:threethree}. In Sec.~\ref{sec:sbnprogram}, we describe the SBN experimental facility in more detail. In Sec.~\ref{sec:sensitivity}, we describe the analysis method followed to estimate SBN's sensitivity to (3+$N$) sterile neutrino oscillations; more specifically, in Sec.~\ref{sec:spectra} we describe the method used to predict the SBN measureable event spectra given any set of (3+$N$) oscillation parameters, and in Sec.~\ref{sec:simulation} we describe the SBN fitting framework and $\chi^2$ calculation method. We present sensitivity results for (3+1), (3+2) and (3+3) in Sec.~\ref{sec:results}, and we further explore SBN's sensitivity to CP-violating phases measurable in the (3+2) scenario in Sec.~\ref{sec:cp}. Finally, a summary and conclusions are provided in Sec.~\ref{sec:conclusion}.

\section{\label{sec:formalism}Sterile neutrino oscillation formalism}

To account for three-neutrino oscillations, the Standard Model prescribes three neutrinos that are pure and distinct eigenstates of the weak interaction: $\nu_e$, $\nu_{\mu}$, and $\nu_{\tau}$, each of which is a linear combination of three distinct neutrino mass eigenstates. The weak eigenstates are defined as
\begin{eqnarray}\label{eq:threeone}
|\nu_\alpha \rangle = \sum^3_{i=1} U^*_{\alpha i} |\nu_i \rangle~,
\end{eqnarray}
where $\alpha=e,\mu$ or $\tau$, and $U_{\alpha i}$ represents the elements of the Pontecorvo-Maki-Nakagawa-Sakata (PMNS) matrix, a $3\times3$, unitary, leptonic mixing matrix.

To determine the probability of a neutrino of flavor $\alpha=e,\mu,\tau$ to be detected as flavor $\beta$ after travelling some distance $L$ and having energy $E$, one may treat the neutrino as a plane wave and evolve the waveform over time. This gives an ``oscillation'' probability of
\begin{eqnarray}
P( & \nu_\alpha &\rightarrow \nu_\beta) = \delta_{\alpha\beta} \nonumber \\
 &-& \sum_{i,j} U^*_{\alpha i}U_{\beta i}U_{\alpha j}U^*_{\beta j}(1 - \exp(i \Delta m^2_{ij} L/2E))~, 
\end{eqnarray}
where $i$ and $j$ run over the three neutrino mass eigenstate indices, and $\Delta m^2_{ij} = m^2_i - m^2_j$ define the mass-squared splitting between any two of the three neutrino mass states. The expression can be further parametrized as
\begin{eqnarray}
P( & \nu_\alpha & \rightarrow \nu_\beta) = \delta_{\alpha\beta} \nonumber \\
&-& \sum_{ij} 4\Re[U^*_{\alpha i}U_{\beta i}U_{\alpha j}U^*_{\beta j}]\sin^2(1.27 \Delta m^2_{ij} L/E) \nonumber \\
&+& \sum_{ij} 2\Im[U^*_{\alpha i}U_{\beta i}U_{\alpha j}U^*_{\beta j}]\sin(2.54 \Delta m^2_{ij} L/E)~,\label{eq:genosc} 
\end{eqnarray}
where we have adopted natural units, $\hbar=c=1$. Antineutrino oscillation can be similarly calculated by substituting the mixing matrix elements $U_{\alpha i}$ with their complex conjugates, $U_{\alpha i}^*$.

From the general oscillation probability formula in Eq.~\ref{eq:genosc}, one can add the effects of $N$ sterile neutrinos by expanding the PNMS matrix to a $(3+$N$)\times(3+N)$, unitary mixing matrix, and summing over $i=1,...,3+N$ distinct mass eigenstates. In this paper, it is assumed that the additional neutrino mass states, $m_4$, $m_5$, and $m_6$, will each be on the order of 0.1-10~eV, which follows from past and recent global fits~\cite{Conrad:2012qt,Collin:2016rao}. The two lowest mass-squared splittings, $\Delta m^2_{21}$ and $\Delta m^2_{32}$, are both well-established through multiple independent experiments and of order $10^{-5}$ eV$^2$ and $10^{-3}$ eV$^2$. As both are sufficiently small, one may apply the short-baseline approximation to this formalism, wherein the three lowest mass states are set to be degenerate at $m_1\sim m_2\sim m_3 \sim 0$~eV. This also assumes a hierarchy where the $\nu_1$, $\nu_2$ and $\nu_3$ mass states are the lightest.

With the above assumptions and approximations, for a (3+1) model, the oscillation probabilities for appearance and disappearance are given by
\begin{align}
\begin{split}
	P(\nu_\alpha\rightarrow\nu_\beta) = 4|U_{\alpha 4}|^2|U_{\beta 4}|^2 \sin^2x_{41},
\end{split}  
\end{align}
and
\begin{align}
\begin{split}
	P(\nu_\alpha\rightarrow\nu_\alpha) = 1 - 4|U_{\alpha 4}|^2&(1 -|U_{\alpha 4}|^2) \sin^2x_{41}~,
\end{split}  
\end{align}
respectively, where $x_{ij}\equiv 1.27\Delta m^2_{ij}L/E$. Thanks to the short-baseline approximation and the unitarity of the PMNS matrix, this case bears striking resemblance to a two neutrino oscillation.

For a (3+2) model, the oscillation probability is given by
\begin{eqnarray}
P(\nu_\alpha \rightarrow \nu_\beta) & = & 4|U_{\alpha4}|^2|U_{\beta4}|^2\sin^2x_{41}  \nonumber \\
    &+& 4|U_{\alpha5}|^2|U_{\beta5}|^2\sin^2x_{51}  \nonumber \\
    &+& 8|U_{\alpha4}||U_{\beta4}||U_{\alpha5}||U_{\beta5}|\times  \nonumber \\
    &~& \sin x_{41}\sin x_{51}\cos(x_{54}-\phi_{54}),
    \label{eq:3p2app}
\end{eqnarray}
in the case of appearance ($\beta\ne\alpha$), and 
\begin{eqnarray}
	P(\nu_\alpha \rightarrow \nu_\alpha) &=& 1 - 4(1-|U_{\alpha 4}|^2-|U_{\alpha 5}|^2)\cdot \nonumber \\
    &~&(|U_{\alpha 4}|^2\sin^2x_{41}+|U_{\alpha 5}|^2\sin^2x_{51})  \nonumber \\
  &-&4|U_{\alpha 4}|^2|U_{\alpha 5}|^2\sin^2x_{54},
\end{eqnarray}
in the case of disappearance. Note that for (3+$N$) neutrino models with N $>$ 1, one must consider the complex phases of the (extended) mixing matrix. Those appear as CP-violating phases $\phi_{ij}$ in the oscillation probability, and are defined as $\phi_{ij} = \arg\{U^*_{\alpha i}U_{\beta i}U_{\alpha j}U^*_{\beta j}\}$ for neutrino oscillations, and $\phi_{ij} = \arg\{U^*_{\beta i}U_{\alpha i}U_{\beta j}U^*_{\alpha j}\}$ for antineutrino oscillations. This is equivalent to substituting $\phi_{ij}$ with $-\phi_{ij}$ in Eq.~\ref{eq:3p2app} when considering {\it antineutrino} appearance probabilities.

Lastly, the (3+3) oscillation probability is given by
\begin{eqnarray}
	P(&\nu_\alpha& \rightarrow \nu_\beta) = \nonumber \\
& -&4|U_{\alpha 5}||U_{\beta 5}||U_{\alpha 4}||U_{\beta 4}| \cos\phi_{54} \sin^2x_{54} \nonumber \\
& -&4|U_{\alpha 6}||U_{\beta 6}||U_{\alpha 4}||U_{\beta 4}| \cos\phi_{64} \sin^2x_{64} \nonumber \\
& -&4|U_{\alpha 5}||U_{\beta 5}||U_{\alpha 6}||U_{\beta 6}| \cos\phi_{65} \sin^2x_{65} \nonumber \\
& +&4(|U_{\alpha 4}||U_{\beta 4}| + |U_{\alpha 5}||U_{\beta 5}| \cos\phi_{54} \nonumber \\
& ~&+|U_{\alpha 6}||U_{\beta 6}| \cos \phi_{64})|U_{\alpha 4}||U_{\beta 4}|\sin^2x_{41} \nonumber \\
& +&4(|U_{\alpha 4}||U_{\beta 4}| \cos \phi_{54} + |U_{\alpha 5}||U_{\beta 5}| \nonumber \\
& ~&+|U_{\alpha 6}||U_{\beta 6}| \cos \phi_{65})|U_{\alpha 5}||U_{\beta 5}|\sin^2x_{51} \nonumber \\
& +&4(|U_{\alpha 4}||U_{\beta 4}| \cos \phi_{64} + |U_{\alpha 5}||U_{\beta 5}| \cos \phi_{65}  \nonumber \\
& ~&+|U_{\alpha 6}||U_{\beta 6}|)|U_{\alpha 6}||U_{\beta 6}|\sin^2x_{61} \nonumber \\
&+&2|U_{\beta 5}||U_{\alpha 5}||U_{\beta 4}||U_{\alpha 4}|\sin \phi_{54} \sin 2x_{54} \nonumber \\
&+&2|U_{\beta 6}||U_{\alpha 6}||U_{\beta 4}||U_{\alpha 4}|\sin \phi_{64} \sin 2x_{64} \nonumber \\
&+&2|U_{\beta 6}||U_{\alpha 6}||U_{\beta 5}||U_{\alpha 5}|\sin \phi_{65} \sin 2x_{65} \nonumber \\
&+&2(|U_{\alpha 5}||U_{\beta 5}|\sin \phi_{54} + |U_{\alpha 6}||U_{\beta 6}|\sin \phi_{64}) \nonumber \\
&~&|U_{\alpha 4}||U_{\beta 4}|\sin 2x_{41} \nonumber \\
&+&2(-|U_{\alpha 4}||U_{\beta 4}|\sin\phi_{54} + |U_{\alpha 6}||U_{\beta 6}|\sin\phi_{65}) \nonumber \\
&~&|U_{\alpha 5}||U_{\beta 5}|\sin 2x_{51} \nonumber \\
&+&2(-|U_{\alpha 4}||U_{\beta 4}|\sin\phi_{64} - |U_{\alpha 4}||U_{\beta 5}|\sin\phi_{65}) \nonumber \\
&~&|U_{\alpha 6}||U_{\beta 6}|\sin 2x_{61},
\end{eqnarray}
in the case of appearance, and 
\begin{eqnarray}\label{eq:threethree}
	P(&\nu_\alpha& \rightarrow \nu_\alpha) = 1-4|U_{\alpha 4}|^2|U_{\alpha 5}|^2\sin^2x_{54}\nonumber \\
    &-&4|U_{\alpha 4}|^2|U_{\alpha 6}|^2\sin^2x_{64}\nonumber \\
    &-&4|U_{\alpha 5}|^2|U_{\alpha 6}|^2\sin^2x_{65}\nonumber \\
    &-&4(1-|U_{\alpha 4}|^2-|U_{\alpha 5}|^2-|U_{\alpha 6}|^2)(|U_{\alpha 4}|^2\sin^2x_{41}\nonumber \\
    &~&+|U_{\alpha 5}|^2\sin^2x_{51}+|U_{\alpha 6}|^2\sin^2x_{61}),
\end{eqnarray}
in the case of disappearance. In this case, there are three CP-violating phases which are free parameters of the model, $\phi_{54}$, $\phi_{64}$ and $\phi_{65}$. 

\section{\label{sec:globalfit}Globally Allowed (3+$N$) Parameter Space}

For any (3+$N$) scenario under consideration, we first perform a fit over existing short-baseline neutrino experiment data, to extract the globally-allowed 90\% and 99\% confidence level (CL) regions over the full available oscillation parameter space. This is done primarily out of computational considerations, in order to obtain a reduced oscillation phase-space over which we subsequently quantify the SBN sensitivity. The data sets included in the global fit are summarized in Tab.~\ref{tab:datasets}, following the methods in Ref.~\cite{Conrad:2012qt}. We omit the recent MINOS+ \cite{MINOS:2016viw} and IceCube~\cite{TheIceCube:2016oqi} constraints from the global fits, although we note that in the future those constraints should be included for more quantitatively accurate results. We expect that the qualitative conclusions drawn in this work stand regardless of inclusion of these more recent constraints in the fit or not.

\begin{table}[tbp]
\begin{tabular}{lc}
\hline \hline 
Dataset & Oscillation Channel  \\ \hline \hline 
Appearance & \\
\hline 
KARMEN~\cite{ds:karmen} & $\numubar \to \bar{\nu}_e$  \\ 
LSND~\cite{ds:lsnd}     & $\numubar \to \bar{\nu}_e$  \\ 
MiniBooNE - BNB~\cite{ds:mbnu,ds:mbnu2,ds:mbnubar,ds:mbnubar2}  & $\overset{(-)}{\nu}_\mu \to \overset{(-)}{\nu}_e$        \\ 
MiniBooNE - NuMI~\cite{ds:numi}          & $\numu \to \nue$      \\ 
NOMAD~\cite{ds:nomad}                     & $\numu \to \nue$    \\ \hline  
Disappearance & \\ \hline 
KARMEN, LSND (xsec)~\cite{ds:xsec}       & $\nue \to \nue$        \\ 
Gallium (GALLEX and SAGE)~\cite{ds:gal,ds:gal2} & $\nue \to \nue$     \\ 
Bugey~\cite{ds:bugey,ds:bugey2}                    & $\nuebar \rightarrow \bar{\nu}_e $   \\ 
MiniBooNE - BNB~\cite{ds:mbnudis,ds:mbnubardis}  & $\overset{(-)}{\nu}_\mu \to \overset{(-)}{\nu}_\mu$                   \\ 
MINOS-CC~\cite{ds:minos, ds:minos2}      & $\numubar \to \bar{\nu}_{\mu}$  \\ 
CCFR84~\cite{ds:ccfr}                    & $\numu \to \numu$   \\ 
CDHS~\cite{ds:cdhs}                      & $\numu \to \numu$    \\ 
Atmospheric Constraints~\cite{ds:superk1,ds:superk2,ds:k2k1,ds:k2k2,ds:k2k3}                      & $\numu \to \numu$   \\ \hline \hline 
\end{tabular}
\caption{\label{tab:datasets}The short-baseline oscillation data sets included in global fits to (3+$N$) sterile neutrino oscillation scenarios, and used to provide allowed regions over which SBN's sensitivity is quantified.}
\end{table}

For each experimental data set included in the global fit, a Monte Carlo prediction is calculated using the oscillation probability derived for a given set of sterile neutrino oscillation parameters and for a given oscillation scenario (Eqs.~\ref{eq:threeone}-\ref{eq:threethree}), and compared against observed data from the experiment. The resulting $\chi^2$ for each experimental data set is summed to form a global $\chi^2$ for each sterile neutrino model, assuming that there are no correlations among the data sets considered in the fit.

Given the broad parameter space in these fits, particularly for the (3+3) scenario that features twelve (12) independent mixing parameters, a grid scan of any reasonable resolution would be very computationally costly. Instead, the scanning of mixing parameters for each oscillation scenario is done more efficiently using a Markov chain \chisq minimization routine, following the method employed in Ref.~\cite{Conrad:2012qt}. The range over which each oscillation fit parameter is defined is set as follows:
\begin{itemize}
\item $0\le U_{\alpha i} \le 0.5$~,
\item $0.01 \le \Delta m^2_{i1} \le 100$~eV$^2$~,
\item $0\le \phi_{ij} < 2\pi$~, 
\end{itemize}
where $\alpha=e,\mu$ and $i,j=4,...,3+N$. Initial values for the $N$ additional neutrino mass states, mixing matrix elements and CP-violating phase(s) are generated randomly from within their corresponding ranges. Then, each of the fit parameters $\theta$ is generated for each successive step in the minimization chain using
\begin{align}
\theta_{new} = \theta_{old} + (R -0.5) (\theta_{max} - \theta_{min}) s~,
\end{align}
where $R$ is a random number in (0,1) and $s$ is a configurable step size scale. Further constraints are applied to all generated $U_{\alpha i}$, consistent with unitarity bounds, by rejecting points in the parameter space where any of the following definitions are invalid:
\begin{itemize}
\item $\sum_{i=4,...,3+N} |U_{\alpha i}|^2 \le 0.3$ for $\alpha = e,\mu$~, or 
\item $\sum_{\alpha=e,\mu} |U_{\alpha i}|^2 \le 0.3$ for $i=4,5,6$~.
\end{itemize}

For each step meeting the above constraints, a (global) $\chi^2$ is calculated for the given set of oscillation parameters $\theta$ by fitting to the experimental data sets. The resulting $\chi^2$ is then compared against the $\chi^2$ calculated in the previous point in the chain, $\chi^2_{old}$, to determine the probability $P_T$ of accepting this new point into the Markov chain. This probability is given by
\begin{align}
P_T = \min(1, exp(-(\chi^2 - \chi^2_{old})/T)),
\end{align}
where $T$ is also a configurable parameter in the Markov chain. By randomly varying the values of $R$, $s$ and $T$, one can combine multiple minimization chains to reach the global minimum $\chi^2$ while evading local minima. 

The resulting global $\chi^2$ multi-dimensional surface is used to determine the parameter space allowed at a certain confidence level, using a $\Delta\chi^2$ cut relative to the global $\chi^2$ minimum, $\chi^2_{min}$. Once a globally-allowed region for a certain scenario is obtained, the region gets discretized over a grid of $100^{n}$ spacepoints, where $n$ is the number of oscillation parameters in the given scenario. The spacepoints are evenly distributed over the ranges defined above, and in a linear scale in mixing elements $U_{\alpha i}$ and a logarithmic scale in $\Delta m^2_{i1}$. Only for the purpose of illustrating two-dimensional projected allowed regions, we marginalize over the oscillation parameter space and thus a $\Delta\chi^2$ cut of 4.61 (90\% CL) and 9.21 (99\% CL) using 2 degrees of freedom ($dof$) is applied. However, to extract the $n$-dimensional phase-space over which we later quantify the SBN sensitivity, the $\Delta\chi^2$ cuts applied more appropriately correspond to $n$ $dof$, where $n=3,7$ and 12 $dof$ for (3+1), (3+2) and (3+3), respectively.

The following subsections provide a summary of the global fit results that are used as input to the SBN sensitivity studies.

\subsection{(3+1) Globally Allowed Regions \label{sec:threeone}}

\begin{figure}[h!]
\begin{center}
  \includegraphics[scale=.15]{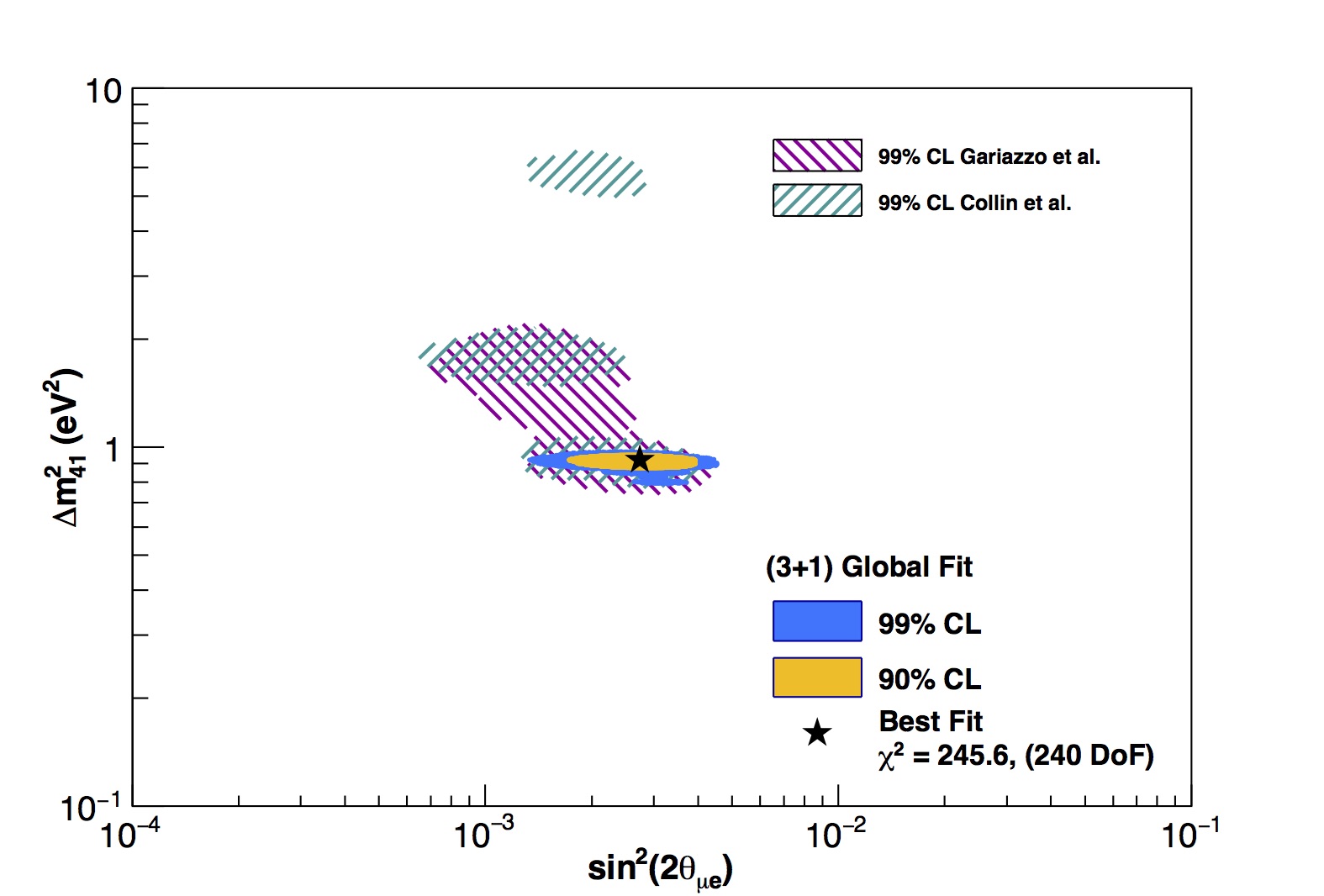}
  \end{center}
  \caption{ The 90\% and 99\% CL regions allowed by a simultaneous fit to all data sets listed in Tab.~\ref{tab:datasets} and following the prescription in Sec.~\ref{sec:globalfit}, under a (3+1) sterile neutrino oscillation hypothesis. Overlaid are results from other recent global fit analyses, including new constraints from the IceCube experiment \cite{TheIceCube:2016oqi}. There are three free oscillation parameters in this fit, but here we profile over them to provide 2D projections in regions of $\Delta m^2_{41}$ and $\sin^2 2\theta_{\mu e}=4|U_{e4}|^2|U_{\mu4}|^2$ that are allowed at the chosen confidence levels, assuming 2~$dof$.}
\label{fig:globfit31}
\end{figure}

In this subsection, we summarize the results of the global fit to all data sets listed in Tab.~\ref{tab:datasets} under the (3+1) oscillation hypothesis. The best fit parameters obtained in this fit, and corresponding $\chi^2_{min}/$$dof$, are provided in Tab.~\ref{bestfits}. A two-dimensional allowed region profiled into $\Delta m^2_{41}$-$\sin^2 2 \theta_{\mu e}$ is illustrated in Fig.~\ref{fig:globfit31}, where $\sin^2 2\theta_{\mu e} = 4|U_{e4}|^2 |U_{\mu4}|^2$. The region at around 1~eV$^2$ is largely driven by the LSND and MiniBooNE anomalies. Note, however, that the recent IceCube constraints tend to shift this allowed region slightly, to higher $\Delta m_{41}^2$ and slightly lower mixing amplitudes. The $\chi^2$ difference between the $\Delta m_{41}^2\sim1$~eV$^2$ and $\Delta m_{41}^2\sim2$~eV$^2$ regions in terms of $\chi^2$ has been reported to be very small, suggesting that one of those new regions is only marginally preferred over the other. For this reason we have chosen to carry out sensitivity studies without the IceCube constraints included for the time being.

\subsection{(3+2) Globally Allowed Regions }

\begin{figure}[ht!]
\begin{center}
  \includegraphics[scale=.15]{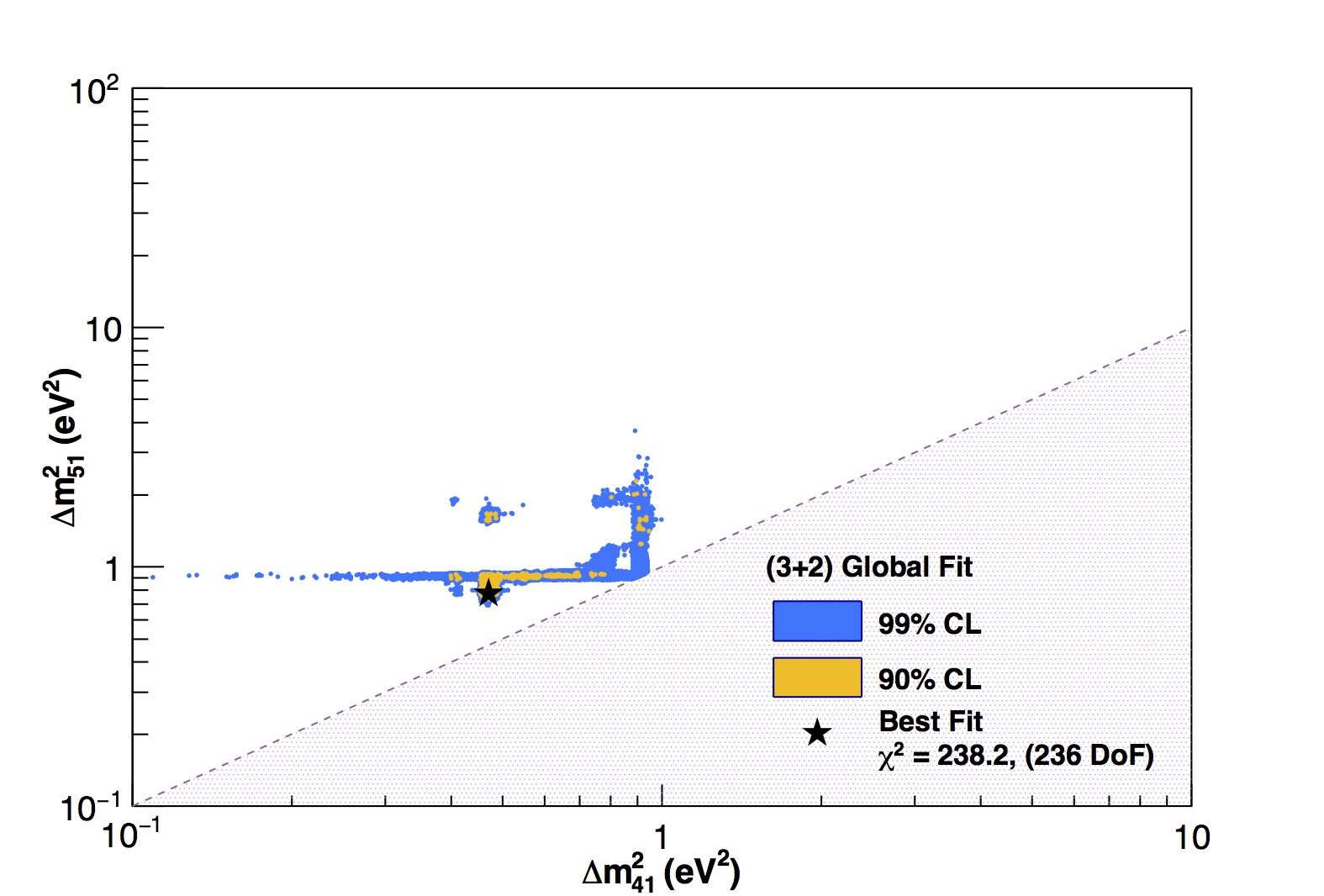}
  \end{center}
  \caption{The 90\% and 99\% CL regions allowed by a simultaneous fit to all data sets listed in Tab.~\ref{tab:datasets}, and following the prescription in Sec.~\ref{sec:globalfit}, under a (3+2) sterile neutrino oscillation hypothesis. There are seven free oscillation parameters in this fit, but here we marginalize over them to provide 2D projections in regions of $\Delta m^2_{41}$ and $\Delta m^2_{51}$ that are allowed at the chosen confidence levels, assuming 2~$dof$.}
  \label{fig:globfit32}
\end{figure}

In this subsection, we summarize the results of the global fit to all data sets listed in Tab.~\ref{tab:datasets} under the (3+2) oscillation hypothesis. The best fit parameters obtained in this fit, and corresponding $\chi^2_{min}/dof$, are provided in Tab.~\ref{bestfits}. A two-dimensional allowed region profiled into ($\Delta m^2_{41}$,$\Delta m^2_{51}$) space is illustrated in Fig.~\ref{fig:globfit32}.

By adding a second light sterile neutrino, one also adds a CP-violating phase, $\phi_{54}$. This additional phase can be influential at short baselines and can relieve some of the tension between neutrino and antineutrino data sets, providing a better overall fit to global data. This improvement has been demonstrated to be the case in particular when considering appearance-only data sets (see, e.g.~\cite{Kopp:2013vaa,Giunti:2015jvi, Conrad:2012qt}). 

\begin{table}[tbp]
\begin{tabular}{|c|ccc|c|}
\hline \hline
(3+1)      & $\Delta m^2_{41}$ & $U_{\mu 4}$ & $U_{e 4}$ & $\chi^2$/$dof$\\ \hline\hline
Best Fit & 0.92                & 0.17          & 0.15    & 245.6/240    \\ \hline
\end{tabular}
\\[.5cm]
\begin{tabular}{|c|ccc|ccc|c|c|}
\hline \hline
(3+2)      & $\Delta m^2_{41}$ & $U_{\mu 4}$ & $U_{e 4}$ & $\Delta m^2_{51}$ & $U_{\mu 5}$ & $U_{e 5}$ & $\phi_{54}$ & $\chi^2$/$dof$ \\ \hline\hline
Best Fit & 0.46                & 0.15          & 0.13        & 0.77                & 0.13          & 0.14        & 5.56   & 238.2/236         \\ \hline \hline 
\end{tabular}
\\[.5cm]
\begin{tabular}{|c|ccc|ccc|ccc|}
 \hline \hline
(3+3)      & $\Delta m^2_{41}$ & $U_{\mu 4}$ & $U_{e 4}$ & $\Delta m^2_{51}$ & $U_{\mu 5}$ & $U_{e 5}$ & $\Delta m^2_{61}$ & $U_{\mu 6}$ & $U_{e 6}$  \\  \hline\hline
Best Fit & 0.68                & 0.18          & 0.12        & 0.90                & 0.13          & 0.14        & 1.55                & 0.03          & 0.12            \\  \hline
  &    $\phi_{54}$ & $\phi_{64}$ & $\phi_{65}$ & $\chi^2$/$dof$ & & & &  & \\ \hline
  &    5.60          & 4.31          & 3.93  & 232.5/231 & & & & &\\ \hline \hline
\end{tabular}
\caption{\label{bestfits}Global best-fit parameters obtained under the (3+1) (top), (3+2) (middle) and (3+3) (bottom) oscillation hypothesis. Mass-squared splittings are presented in eV$^2$ and CP-violating factors are given in radians. The null hypothesis has a $\chi^2$/$dof=299.5$/$243$. }
\end{table}

\begin{figure}[b!]
\begin{center}
  \includegraphics[scale=.15]{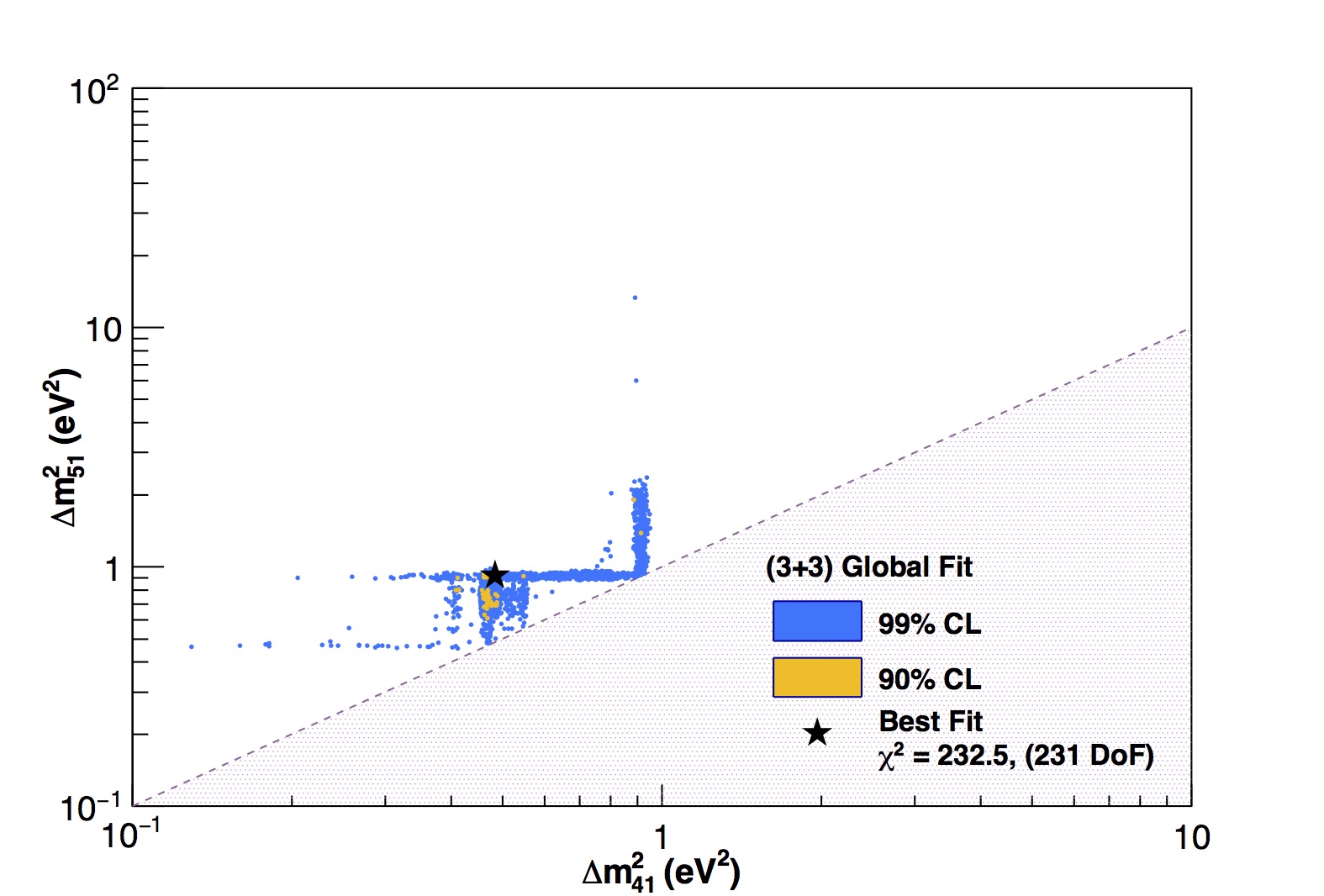}
  \caption{The 90\% and 99\% CL regions allowed by a simultaneous fit to all data sets listed in Tab.~\ref{tab:datasets}, and following the prescription in Sec.~\ref{sec:globalfit}, under a (3+3) sterile neutrino oscillation hypothesis. There are twelve free oscillation parameters in this fit, but here we marginalize over them to provide 2D projections in regions of $\Delta m^2_{41}$ and $\Delta m^2_{51}$ that are allowed at the chosen confidence levels, assuming 2~$dof$.}
  \label{fig:globfit33.1}
  \end{center}
\end{figure}

\begin{figure}[ht!]
\begin{center}
  \includegraphics[scale=.15]{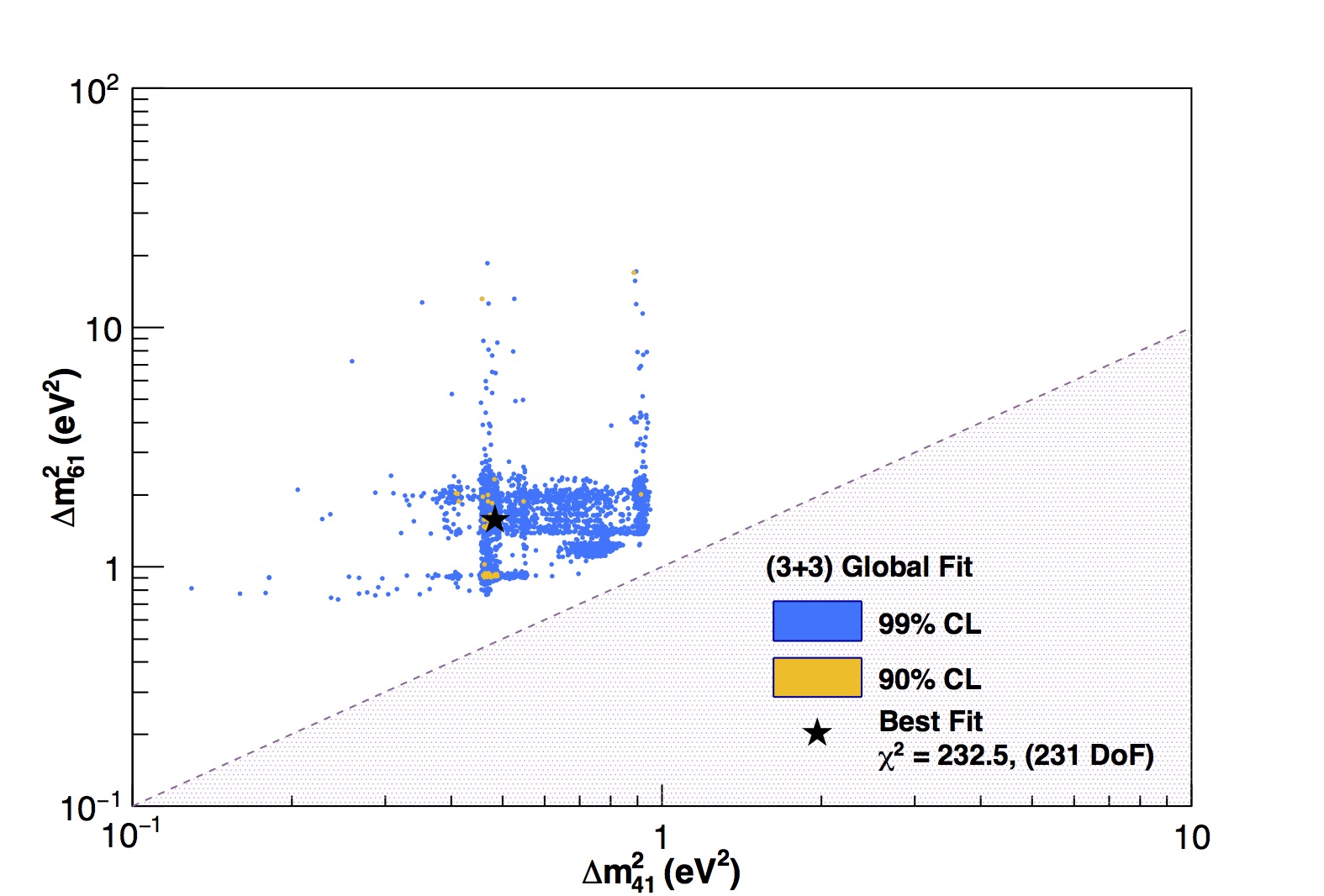}
  \caption{The 90\% and 99\% CL regions allowed by a simultaneous fit to all data sets listed in Tab.~\ref{tab:datasets}, and following the prescription in Sec.~\ref{sec:globalfit}, under a (3+3) sterile neutrino oscillation hypothesis. There are twelve free oscillation parameters in this fit, but here we marginalize over them to provide 2D projections in regions of $\Delta m^2_{41}$ and $\Delta m^2_{61}$ that are allowed at the chosen confidence levels, assuming 2~$dof$.}
  \label{fig:globfit33.2}
  \end{center}
\end{figure}

\subsection{(3+3) Globally Allowed Regions \label{sec:threethree}}

In this subsection, we summarize the results of the global fit to all data sets listed in Tab.~\ref{tab:datasets} under the (3+3) oscillation hypothesis. The best fit parameters obtained in this fit, and corresponding $\chi^2_{min}/dof$, are provided in Tab.~\ref{bestfits}. Two-dimensional allowed regions profiled into ($\Delta m^2_{41}$,$\Delta m^2_{51}$) and ($\Delta m^2_{51}$,$\Delta m^2_{61}$) space are illustrated in Figs.~\ref{fig:globfit33.1} and \ref{fig:globfit33.2}, respectively.

The addition of yet another light sterile degree of freedom comes with five additional parameters, including an additional independent mass splitting, two additional mixing elements, and two additional CP-violating phases. This further increases the hypervolume of parameter space allowed under the global data sets, although the preference for one of the best fit $\Delta m^2_{i1}$ being close to $\mathcal{O}(1\text{eV}^2)$ evident in the (3+1) and (3+2) hypotheses seems to persist. Furthermore, as in the (3+2) case, the additional CP-violating phases in the (3+3) case have been shown to lead to a further reduction in tension between neutrino and antineutrino data sets as well as an overall lessening of the disagreement between appearance-only and disappearance-only fits (see, e.g., Refs.~\cite{Kopp:2013vaa,Giunti:2015jvi, Conrad:2012qt}). 

\section{\label{sec:sbn}SBN Sensitivity to (3+$N$) Oscillations}

\subsection{\label{sec:sbnprogram}The SBN Program}

The Short Baseline Neutrino (SBN) program aims to perform a highly sensitive search for sterile neutrino oscillations at an $L/E$ of $\sim 1$~km/GeV. The program utilizes three LArTPC detectors---ICARUS, MicroBooNE and SBND---each placed at a different baseline $L$ along the Booster Neutrino Beam (BNB) line at Fermilab.
  
ICARUS is the first large-scale LArTPC neutrino detector ever constructed, and has previously operated at the Gran Sasso National Laboratory in Italy. It is presently being refurbished and prepared for transit to Fermilab in Spring of 2017. It has an active mass of 476 tons of liquid argon and will be placed 600 meters from neutrino production in the BNB, forming the far detector of the SBN program. MicroBooNE is the mid detector, and it has already begun operations in the BNB, in October 2015. The MicroBooNE active mass is 89 tons, and the detector is located at 470 meters from neutrino production, at roughly the same baseline as its predecessor MiniBooNE experiment. MicroBooNE is on track to collect data corresponding to a beam delivery of 6.6e20 protons on target (POT) before concurrent running with SBND and ICARUS begins. SBND will act as a near detector for the SBN program, located at 110 meters from neutrino production and with an active mass of 112 tons. It is currently under construction and is scheduled to begin taking data with ICARUS and MicroBooNE in late 2018~\cite{Antonello:2015lea}.

The strength of the SBN program comes from the utilization of each of these three detectors in concert, sharing the same beam and the same neutrino interaction target (argon). SBND in particular will be recording very high statistics of interactions of the (mostly un-oscillated) neutrino flux, and thus will be capable of constraining flux and cross section systematic uncertainties for the event rate measurements at the farther detectors. Since all three detectors share the same detector technology, their detector systematics are also expected to be correlated to a certain extent. This will grant unprecedented sensitivity to short-baseline neutrino oscillations, allowing for the verification or ruling out of a large area of parameter space for (3+$N$) sterile neutrino oscillations.

\subsection{\label{sec:sensitivity}Sensitivity Analysis Method}

In order to evaluate SBN's sensitivity to ($3+N$) sterile neutrino oscillations, we consider the oscillation-induced fluctuations that are measurable in the exclusive $\nu_e$ (and $\bar{\nu}_e$) and $\nu_{\mu}$ (and $\bar{\nu}_\mu$) charged-current (CC) event spectra of each of the SBN detectors\footnote{Since the detectors are not capable of classifying a single event as either a neutrino or an antinuetrino interaction, we treat reconstructed neutrino and antineutrino events in these spectra indistinguishably.}. The event spectra are provided in terms of reconstructed neutrino energy, and were estimated as described in Sec.~\ref{sec:spectra}. 

The $\nu_e$ CC spectrum at each detector location is sensitive to potential $\nu_{\mu}\rightarrow\nu_{e}$ appearance in the $\nu_{\mu}$-dominated BNB. For this sample, because background contributions are comparable to signal contributions for most of the globally-allowed (3+$N$) oscillation parameter space, we additionally consider the effects of (1) disappearance of the $\nu_e$ intrinsic background in the beam; and (2) disappearance of the mis-identified background from $\nu_{\mu}$ CC interactions. We assume that the mis-identified background from neutral-current (NC) interactions will be measured and constrained independently and in situ for each of the SBN detectors, and therefore we ignore any oscillation variations on that particular background in these fits. 

The $\nu_{\mu}$ CC spectrum, on the other hand, is sensitive to exclusively $\nu_{\mu}$ disappearance. In this case, we ignore not only oscillation variations on any backgrounds, but also background contributions from NC $\pi^\pm$ production events altogether. Based on Ref.~\cite{Antonello:2015lea}, this background contribution has negligible effect on the SBN sensitivity.

Combining $\nu_e$ and $\nu_{\mu}$ CC measurements, and accounting for correlations due to flux and cross-section between the different exclusive samples ($\nu_e$ CC, $\nu_{\mu}$ CC, etc.), baselines (near, mid, far), and beam running mode (neutrino or antineutrino), allows one to simultaneously constrain both appearance and disappearance probabilities for $\nu_e$ and $\nu_{\mu}$ oscillations. We have developed and followed a fit method that allows for these correlations to be exploited, and which also allows for studying these effects in combination or separately. The fit method is described in detail Sec.~\ref{sec:simulation}.

\subsection{\label{sec:spectra}Predicting SBN Event Spectra}

The SBN $\nu_e$ and $\nu_{\mu}$ CC event spectra used in this work were fully simulated on an event-by-event basis. The raw rates of each flavor of neutrino impinging on the three SBN detectors were evaluated using the flux predictions in \cite{LAr1ND_prop}. Events were generated in GENIE 2.8.6 (default settings used) separately for each neutrino type ($\nu_e$, $\nu_{\mu}$, $\bar{\nu}_{e}$, $\bar{\nu}_{\mu}$) and for the beam polarity in both neutrino and antineutrino mode.

Ten million events were generated for each flavor, detector, and beam polarity.
This corresponds to 8e20~POT for the SBND neutrino mode $\nu_{\mu}$ flux, and significantly more for all other samples. Weights were applied to all events to normalize them to the rates predicted by GENIE for the expected exposure and for each detector active mass. The beam exposure assumed for neutrino running mode is the nominal 6.6e20~POT for which the SBN program has been approved to run, plus the preceding 6.6e20~POT with MicroBooNE-only running.

Subsequently to event generation, events were processed further to emulate the reconstruction and selection of $\nu_e$ CC and $\nu_{\mu}$ CC events, following the assumptions provided in \cite{Antonello:2015lea}. More specifically, to estimate detector effects without the need for a full detector simulation, neutrino interaction final state energies were smeared according to a Gaussian around their true value, using the detector energy resolution quoted in \cite{Antonello:2015lea}: 15\%/$\sqrt{E}$ for electrons and photons, and 6\%/$\sqrt{E}$ for muons and pions; all protons with true kinetic energy below 21~MeV were assumed to be non-reconstructable, while those above this threshold as well as other charged hadrons had their kinetic energies smeared by 5\%. All smeared hadronic energies were added to form the hadronic activity, and the reconstructed neutrino energy was then defined as the total sum of visible (smeared) lepton or photon energy and hadronic activity, as well as the rest masses of all leptons and non-proton charged hadrons. A lower threshold of 100~MeV was also placed on electron and photon energies in order for them to be defined as reconstructable, in line with the SBN proposal assumptions. 

The fiducial volume cut efficiency for each detector was then emulated by randomizing the neutrino interaction vertex position within the predefined active detector volumes, and applying geometric cuts, with the position and direction of all final state muons and e/$\gamma$ showers in the simulation accounted for to accurately estimate backgrounds and efficiencies. This is of utmost importance to the $\nu_e$ appearance signal as $\pi^0 \rightarrow \gamma \gamma$ decays, in which only one photon is reconstructed successfully, can be a non-negligible background. 

The following contributions were included explicitly in the $\nu_e$ CC sample:
\begin{itemize}
\item Intrinsic and signal $\nu_e$ CC events: These events are the largest contribution to the $\nu_e$ CC sample. All appearance signal (from potential $\nu_{\mu}\rightarrow\nu_e$ oscillations) and intrinsic beam $\nu_e$ CC events producing an electron with reconstructed neutrino energy $E_\text{reco} \geq$ 200~MeV were included with an overall 80\% identification efficiency. 
\item NC single photon events, from either NC $\Delta$ production followed by radiative decay, or $\pi^0$ production followed by decay into two photons where only one photon is reconstructable, are also considered as a potential background contribution in the $\nu_e$ CC sample. In particular, events in which the photon is reconstructed too close (within 3~cm) to a vertex identified by significant hadronic activity (defined as $E_\text{visible hadronic} \geq 50$ MeV), or in which no hadronic activity is visible, were included as backgrounds if the reconstructed event energy satisfies the 200~MeV threshold. Those selected events received an additional reduction factor scaling assuming a 94\% photon rejection efficiency.
\item $\nu_\mu$ CC events in which the muon is mis-identified as a pion and simultaneously an additional photon (e.g from $\pi^0$ decay) mimics the electron from a $\nu_e$ CC event were also included as a background contribution to the $\nu_e$ CC sample. To quantify this background, all $\nu_{\mu}$ CC events with a track of length $\geq1$~m were assumed to be identifiable as $\nu_{\mu}$-induced CC events and were rejected. Those with a track length below 1~m were accepted as potential mis-identified events, if any photons in the event were accepted under the same conditions as in the NC single photon events, above. 
\item Interactions outside of the TPC producing photons that propagate inside the active volume are a source of background as well. These ``Dirt'' backgrounds were included with rates (per POT) taken directly from Ref.~\cite{Antonello:2015lea}. We assume that independent measurements of these backgrounds at each detector location will render this contribution insensitive to any oscillation effects. 
\item Cosmogenic backgrounds are expected to be well constrained by topological, calometric and timing cuts, with the background contribution scaling linearly with POT. The numbers we use were taken directly from Ref.~\cite{Antonello:2015lea} and correspond to 146, 88 and 164 cosmogenic background events for SBND, MicroBooNE and ICARUS, respectively, for an exposure corresponding to 6.6e20~POT. Although significantly smaller than the intrinsic $\nu_e$ CC backgrounds, they tend to accumulate at low energy, and thus they were included in our analysis following the approach in Ref.~\cite{Antonello:2015lea}.
\end{itemize}

Cosmogenic and dirt background contributions in antineutrino running mode, are taken to be identical (in rate) to the neutrino running mode samples, scaled only according to POT. 

Similarly, for the $\nu_\mu$ CC sample, intrinsic beam $\nu_\mu$ CC events were assumed to be selected with an 80\% reconstruction and identification efficiency. Potential background contributions would result from NC $\pi^\pm$ interactions where the $\pi^\pm$ can be mis-identified as a muon. This background was mitigated by requiring that all contained muon-like tracks have a track length larger than 50~cm, and that all escaping tracks that have a track length of less than 1~m are rejected. This is the same methodology as what was followed in Ref.~\cite{Antonello:2015lea}.

We show our simulated neutrino mode predictions of the $\nu_e$ CC and $\nu_\mu$ CC spectra for the MicroBooNE detector in Fig.~\ref{fig:sample_spec}, along with an estimated appearance-only signal prediction for a benchmark (3+1) sterile neutrino oscillation model with $\Delta m^2_{41} = 1$~eV$^2$ and $\nu_e$ appearance amplitude of $\sin^2 2\theta_{\mu e} = 10^{-3}$ in the upper figure, and $\nu_\mu$ disappearance amplitude of $\sin^2 2 \theta_{\mu \mu} = 0.1$ in the lower figure. The spectra are in reasonable agreement with those provided in Ref.~\cite{Antonello:2015lea}.

\subsection{SBN $\chi^2$ Calculation\label{sec:simulation}}

To facilitate a multi-baseline, multi-channel, and multi-mode (neutrino and potential antineutrino running) oscillation search with the SBN detectors, we use a custom fitting framework to simultaneously fit the reconstructed $\nu_e$ CC and $\nu_{\mu}$ CC inclusive spectra expected at each detector with and without oscillations, and for each running mode, simultaneously. This simultaneous, side-by-side fit of multiple event samples by way of a full covariance matrix that contains statistical and systematic uncertainties as well as systematic correlations among the different samples, baselines, and running modes, builds on a general approach that has been followed by the MiniBooNE collaboration for several analyses, e.g.~\cite{ds:mbnu,ds:mbnu2,ds:mbnubar,ds:mbnubar2,ds:mbnubardis,ds:mbnudis}, as well as by the SBN collaboration. However, this is the first time that multi-channel and multi-mode first are attempted for SBN. We have chosen this approach specifically so that we may exploit powerful correlations shared within and among the spectra that are measurable by each of the three detectors, with the aim of providing stronger constraints to the multi-parameter oscillation hypotheses under consideration.

\begin{figure}[t!]\label{fig:ubspectrum}
\begin{center}
\vspace{0.2cm}
  \includegraphics[width=0.4 \textwidth]{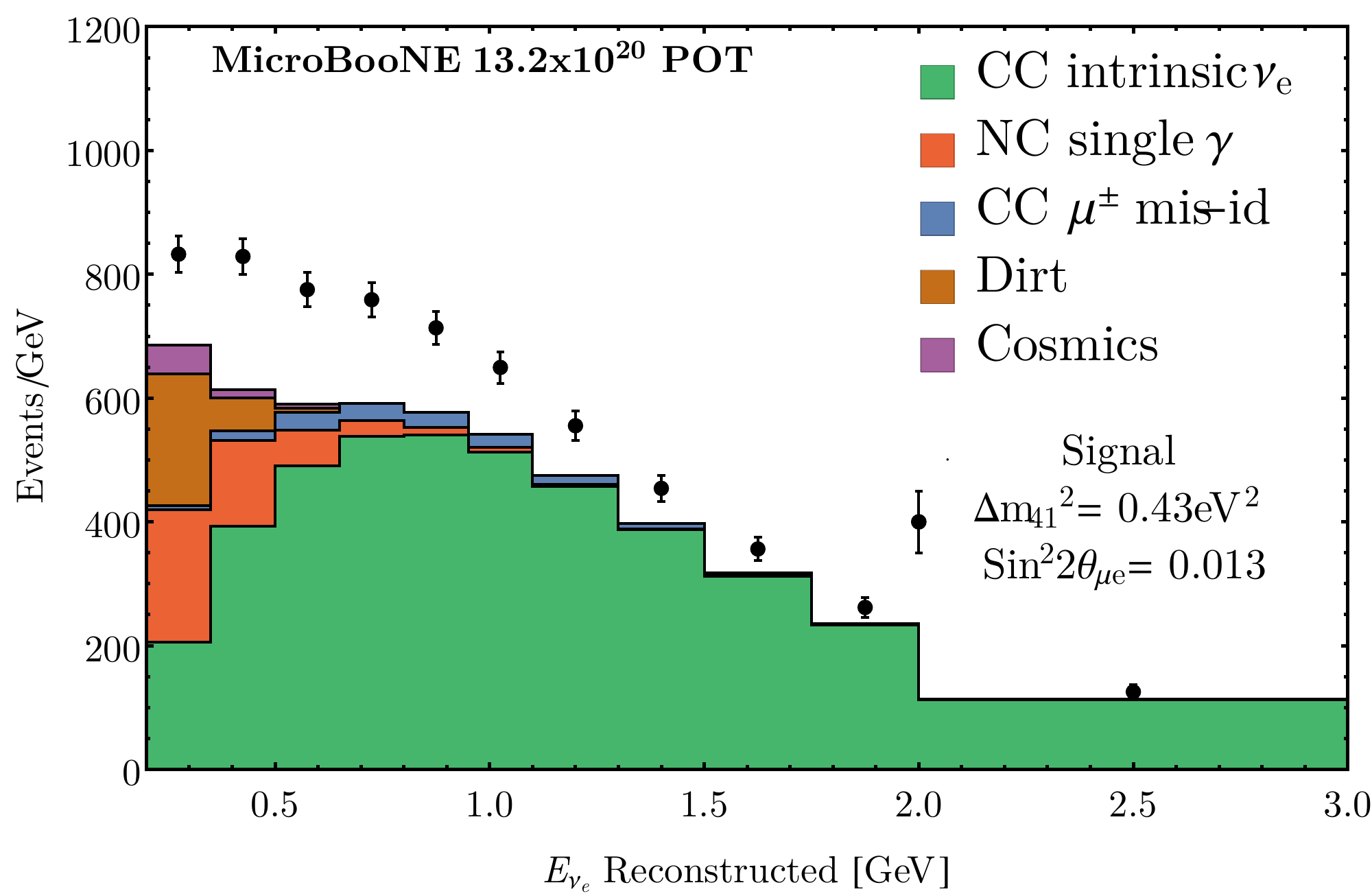}\\
  \hspace{-0.18cm}\includegraphics[width=0.415 \textwidth]{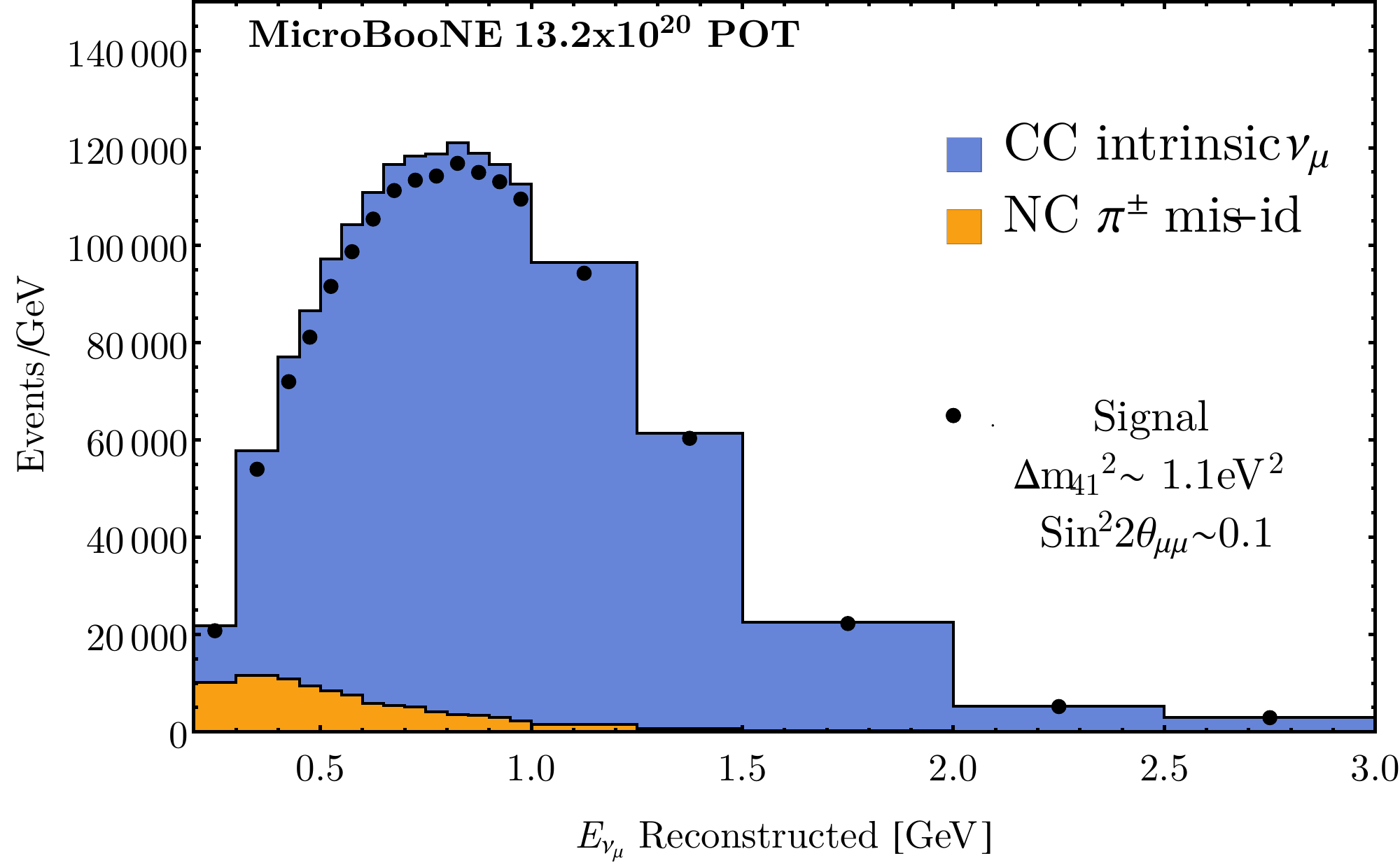}
 
  \end{center}
  \caption{ \emph{Top:} The $\nu_e$ CC inclusive sample used in SBN sensitivity studies, shown only for the MicroBooNE detector. Expected intrinsic and mis-identified backgrounds to $\nu_e$ appearance/disappearance are shown in stacked, colored histograms. Shown also is the expected signal for a benchmark sterile neutrino oscillation model with $\Delta m^2_{41} = 0.43 \text{eV}^2$ and $\sin^2 2\theta_{e \mu} = 0.013$, for comparison. \emph{Bottom:} The $\nu_\mu$ CC inclusive sample used in SBN sensitivity studies, shown only for the MicroBooNE detector, with example $\Delta m^2_{41} = 1.1 \text{eV}^2$ and $\sin^2 2\theta_{\mu \mu} = 0.1$. }
\label{fig:sample_spec}
\end{figure}

The SBN fit quality is quantified over an $n$-dimensional oscillation parameter space volume $(\Delta m^2_{i1},U_{\alpha i},\phi_{ij})$ by way of a $\chi^2$. The $\chi^2$ is calculated over concatenated $\nu_e$ CC inclusive and $\nu_{\mu}$ CC inclusive spectra for all three detectors, as
\begin{widetext}
\begin{equation}
\chi^2(\Delta m^2_{i1},U_{\alpha i},\phi_{ij})=\sum_{k=1}^{M}\sum_{l=1}^{M} \left[ N^\text{null}_k - N^\text{osc}_k(\Delta m^2_{i1},U_{\alpha i},\phi_{ij}) \right] E_{kl}^{-1} \left[ N^\text{null}_l - N^\text{osc}_l(\Delta m^2_{i1},U_{\alpha i},\phi_{ij}) \right]~,
\label{eq:chi}
\end{equation}
\end{widetext}
where $N^{\text{null}}_k$ is the number of events expected under the no oscillation hypothesis (defined as $U_{\alpha i}=0 \quad \forall \quad \alpha,i,j$) in the $k^\text{th}$ bin of reconstructed neutrino energy; $N^{\text{osc}}_k (\Delta m^2_{i1},U_{\alpha i},\phi_{ij})$ is the number of events predicted to be observed in reconstructed neutrino energy bin $k$ under an oscillation hypothesis described by the set of parameter values $(\Delta m^2_{i1},U_{\alpha i},\phi_{ij})$; and $E_{kl}$ is a full $M\times M$ covariance matrix containing the total systematic and statistical uncertainty, including systematic correlations between any two bins $k$ and $l$. The $\nu_e$ CC and $\nu_\mu$ CC samples for each detector location are binned in 11 and 19 bins of reconstructed neutrino energy, respectively, as shown in Fig.~\ref{fig:sample_spec}. Thus, for all three detector locations, the concatenated spectra $N^{\text{null}}_k$ and $N^{\text{osc}}_k$ consist of a total of $M=90$ bins for neutrino-only fits, and $M=180$ bins for neutrino and antineutrino combined fits. 

The covariance matrix, which is a $90\times 90$ matrix for neutrino-only fits, and a $180\times180$ matrix for combined neutrino and antineutrino fits, is calculated as the sum of covariance matrices estimated for each (independent) source of systematic and statistical uncertainty, 
\begin{eqnarray}
E &=& E^\text{stat} + E^\text{flux}+ E^{\text{cross section}} \nonumber \\
&+& E^\text{cosmic}+E^\text{dirt}+E^\text{detector}~.
\end{eqnarray}
Table~\ref{tab:systematics} summarizes the assumed variations on specific contributions to the inclusive $\nu_e$ and $\nu_\mu$ CC samples due to different sources of systematic uncertainty; those variations were used to calculate each corresponding fractional systematics covariance matrix. The assumed numbers are based on Ref.~\cite{Antonello:2015lea}. More specifically, flux systematic uncertainties were estimated by assuming an overall 20\% normalization uncertainty fully correlated among the intrinsic $\nu_e$ (background and signal) and $\nu_\mu$ events, with the exception of exclusive samples that are assumed to be constrained in situ; namely, dirt, cosmogenic, and NC backgrounds in the $\nu_e$ CC sample. A 60\% $\nu_e-\nu_\mu$ flux correlation coefficient was assumed among $\nu_e$ and $\nu_\mu$ events. Cross section systematic uncertainties were estimated by assuming an overall 20\% normalization uncertainty fully correlated among CC-only events, and a corresponding 30\% normalization uncertainty among NC-only events. Again, dirt, cosmogenic, and NC backgrounds in the $\nu_e$ CC sample are exempted from this uncertainty. A 50\% CC-NC cross section correlation coefficient is assumed among CC and NC events. Furthermore, neutrino and antineutrino spectra CC cross section uncertainties are assumed to be 100\% correlated, and likewise for NC cross-section uncertainties. Detector systematics are assumed to be fully uncorrelated among different detectors, and contribute to the overall uncertainty at the level of 2.5\%. These are taken to be fully correlated for neutrino and antineutrino spectra in any given detector. 

\begin{table}
[tb!]
\begin{tabular}{ll}
\hline \hline
Source of Uncertainty & Assumed variation \\ \hline \hline
$\nu_e$ flux & 15.3\% on $\nu_e$ events \\
$\nu_\mu$ flux & 15.1\% on $\nu_\mu$ events \\
CC cross section & 20\% on CC events \\
NC cross section & 30\% on NC events \\
detector effects & 2.5\% on all events \\
\hline \hline
\end{tabular}
\caption{\label{tab:systematics}Assumed variations on exclusive event samples due to different systematic uncertainties, used to evaluate the total systematics covariance matrix. See text for more details.}
\end{table}

The dirt event rate uncertainty is assumed to be constrained through in situ dirt-enhanced sample measurements at each detector and in each running mode. A 15\% normalization uncertainty is assumed for dirt events, taken to be uncorrelated between the different detectors and the neutrino and antineutrino run samples. Similarly, the cosmogenic background uncertainty is assumed to be constrained through in situ off-beam high-statistics rate measurements at each detector. A 1\% normalization uncertainty is assumed for cosmic backgrounds, assumed to be uncorrelated between different detectors, but fully correlated between neutrino and antineutrino samples within any given detector. Finally, NC backgrounds are also assumed to be constrained through an situ NC $\pi^0$ event rate measurements in each detector, thus the estimated statistical uncertainty of the in situ measurement is taken as the systematic uncertainty on these backgrounds. This corresponds to a 0.24\%, 1.3\%, and 5\% normalization uncertainty for the SBND, MicroBooNE, and ICARUS NC background rates, respectively, for 6.6e20~POT. This systematic uncertainty is assumed to be uncorrelated for neutrino and antineutrino run samples. 

When quantifying SBN's sensitivity, we are interested primarily in two fitting methods: 
\begin{itemize}
\item $\nu_e$ appearance-only fits, where $N^{\textrm{osc}}_k(\Delta m^2_{i1},U_{\alpha i},\phi_{ij})$ is evaluated assuming only $\nu_\mu\rightarrow\nu_e$ oscillations, and no $\nu_e$ or $\nu_\mu$ disappearance; this is the method followed by past MiniBooNE oscillation searches \cite{ds:mbnubar} as well as in Ref.~\cite{Antonello:2015lea}; and 
\item combined $\nu_e$ dis/appearance and $\nu_\mu$ disappearance fits, where $N^{\textrm{osc}}_k(\Delta m^2_{i1},U_{\alpha i},\phi_{ij})$ is evaluated assuming $\nu_\mu\rightarrow\nu_e$ oscillations, $\nu_e$ disappearance, as well as $\nu_\mu$ disappearance. We note that this is the first time that SBN sensitivities are evaluated without the implicit assumption of no significant $\nu_e$ or $\nu_\mu$ disappearance; as demonstrated in the results section, this implicit assumption can have a non-negligible effect on the SBN sensitivity.
\end{itemize}

\section{\label{sec:results}SBN Sensitivity to Sterile Neutrino Oscillations: Results}

\begin{figure}[t!]
  \centering
  \includegraphics[width=0.45 \textwidth]{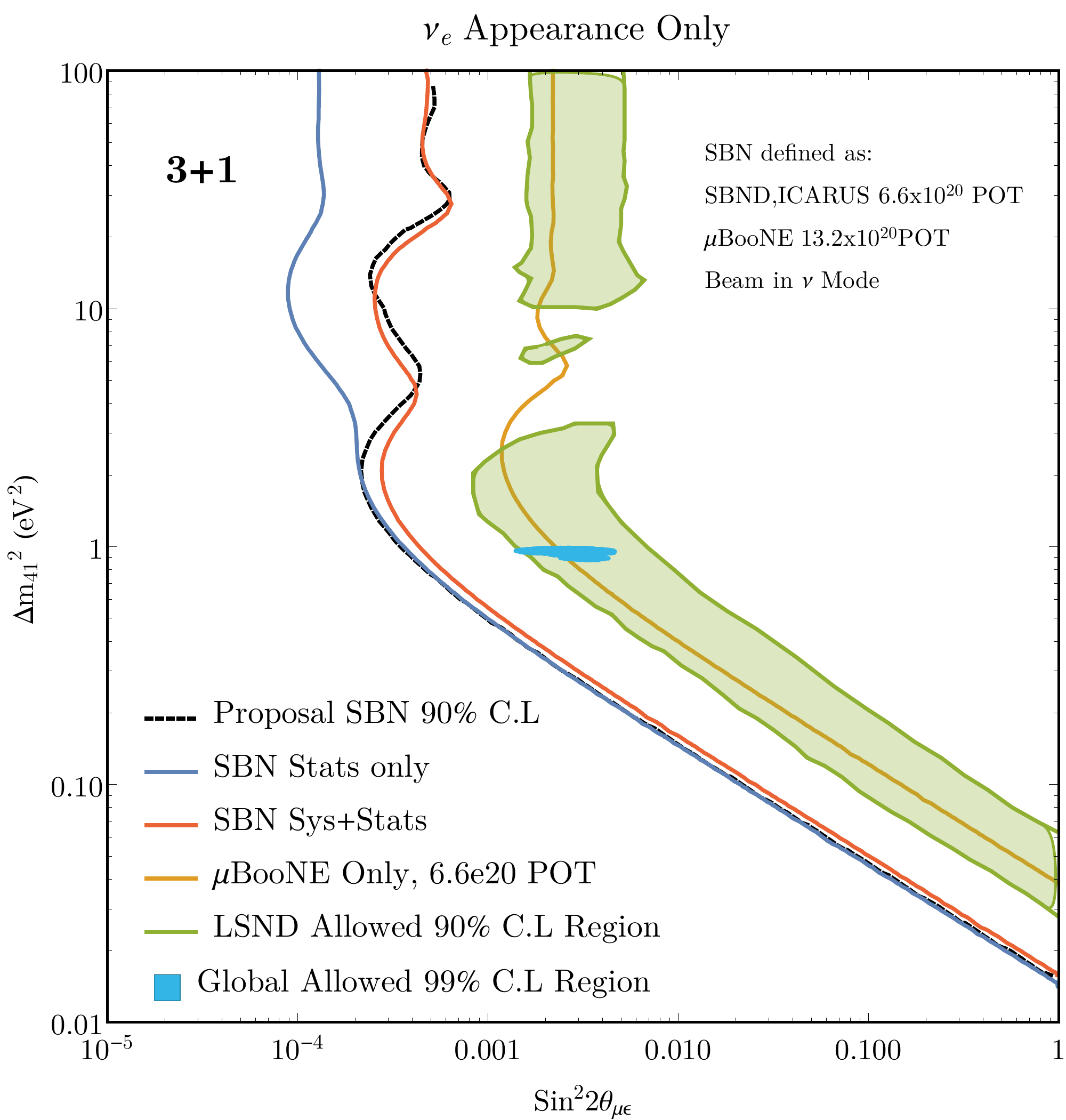}
  \caption{The estimated 90\% CL exclusion contours for the entire SBN program for $\nu_e$ appearance only (yellow solid line) with full detector, flux and cross-section systematics included as well as statistic only (blue).  The same contour as estimated in the SBN proposal is shown in (black dashed) line. This vastly covers the current 99\% (3+1) allowed regions (blue shaded region) and LSND 90\% allowed region (green). Shown also is the $\mu$BooNE only contour (orange) which can probe a large fraction of the global allowed region.}
  \label{fig:3p1_app_only}
\end{figure}

	\subsection{\label{sec:results3p1}(3+1) Scenario at SBN}
Throughout this analysis we will use the globally allowed regions of sterile neutrino parameter space, as described in Sec.~\ref{sec:globalfit}, to investigate what fraction of that parameter space SBN should be able to probe. 

For reference, we first explore SBN's sensitivity reach in neutrino running mode under three separate oscillation assumptions:\\
\begin{itemize}
\item {$\nu_{\mu}\rightarrow\nu_e$ appearance-only (assuming no $\nu_{\mu}$ or $\nu_e$ disappearance). We note that this case involves an odd assumption in a (3+1) oscillation hypothesis, as $\nu_{\mu}\rightarrow\nu_e$ appearance implies both $\nu_{\mu}$ and $\nu_e$ disappearance. However, in the past this case has been applied to MiniBooNE searches to a reasonably good approximation, and has furthermore been applied to SBN sensitivity studies in \cite{Antonello:2015lea}. We therefore consider it only as an instructive example, and to further argue that it is not a reasonable approximation to use for SBN. }
\item { $\nu_{\mu}$ disappearance-only (assuming no $\nu_{e}$ dis/appearance). We consider this case only as an instructive scenario, as the interpretation of short-baseline positive signals also require $\nu_e$ dis/appearance.}
\item { $\nu_{e}$ disappearance-only (assuming no $\nu_{\mu}$ disappearance or $\nu_e$ appearance). We also consider this case only as an instructive scenario, as the interpretation of short-baseline positive signals require both $\nu_e$ and $\nu_\mu$ disappearance (and $\nu_e$ appearance).}
\end{itemize}

Figure~\ref{fig:3p1_app_only} shows the SBN appearance-only sensitivity reach in $\Delta m^2_{41}$ vs.~$\sin^2 2 \theta_{\mu e}$ space under a (3+1) hypothesis obtained using the $\chi^2$ definition described in Sec.~\ref{sec:simulation} and applying a ``raster scan'' over this reduced two-dimensional parameter space. The appearance-only sensitivity is provided here merely for comparison to the sensitivity presented in the SBN proposal \cite{Antonello:2015lea}, which uses the same assumption of no background disappearance, as a means of validating our methodology. The resulting sensitivity in  this work, when incorporating full detector, cross-section and flux systematics (yellow curve), is consistent with the one published in the SBN proposal (black curve). 

The statistics-only sensitivity curve obtained in this work is also shown, in blue. Comparing the blue and red curves demonstrates the effect of systematic uncertainties on the sensitivity, which is to diminish sensitivity to higher-$\Delta m^2_{41}$ oscillations. This is due to the fact that the dominant systematic is the flux and cross-section normalization uncertainty. The comparison also demonstrates the power of exploiting correlations that exist among multiple baselines and multiple interaction channels. Accounting for these correlations leads to an effective cancellation of systematic uncertainties across the three-detector spectra, evident in particular in the low-$\Delta m^2_{41}$ region. Shown also is our projected MicroBooNE-only result after its first run, corresponding 6.6e20~POT. Overlaid over all these curves is the LSND 90\% CL allowed region (shaded green area) as well as the (3+1)  99\% CL globally allowed region from Fig.~\ref{fig:globfit31}. The raster scan sensitivities are obtained using a one-sided $\Delta\chi^2$ cut for 1~$dof$, while the globally allowed region corresponds to a global scan using a $\Delta\chi^2$ cut for 2~$dof$.

\begin{figure}[h!]
  \centering
  \includegraphics[width=0.45 \textwidth]{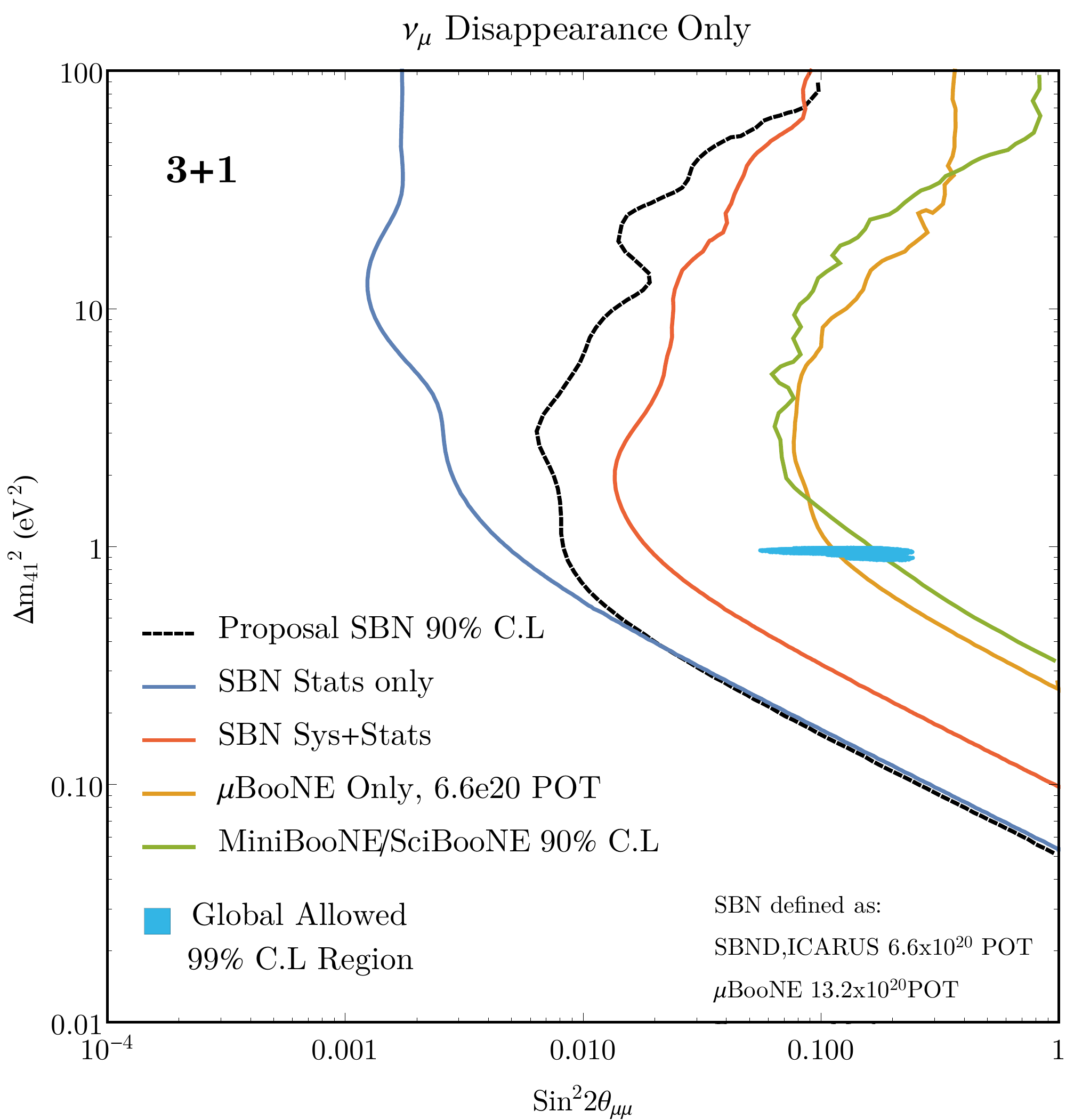}
  \caption{ The estimated 90\% CL contours for the combined SBN using $\nu_\mu$ disappearance only. The globally allowed region in $\Delta m^2_{41}$ and $\sin^2 2 \theta_{\mu \mu}$ is completely covered. Shown also is the prediction for MicroBooNE after 6.6e20~POT.}
  \label{fig:3p1_dis_only}
\end{figure}

The SBN $\nu_\mu$ disappearance-only search gives the sensitivity curve shown in Fig.~\ref{fig:3p1_dis_only} (red curve). As the sensitivity presented in the SBN proposal (black curve) does not include detector systematics, it outperforms the one obtained in this work. This is expected, as detector systematics across the three detectors are taken to be fully uncorrelated in our fits. As a cross check, we compare to the statistics-only sensitivity obtained in this work (blue curve), which is found to lie mostly to the left of both other curves, also as expected. Shown also is the prediction for MicroBooNE ($\mu$BooNE) after its first 6.6e20~POT exposure. 

\begin{figure}[h!]
  \centering
  \includegraphics[width=0.45 \textwidth]{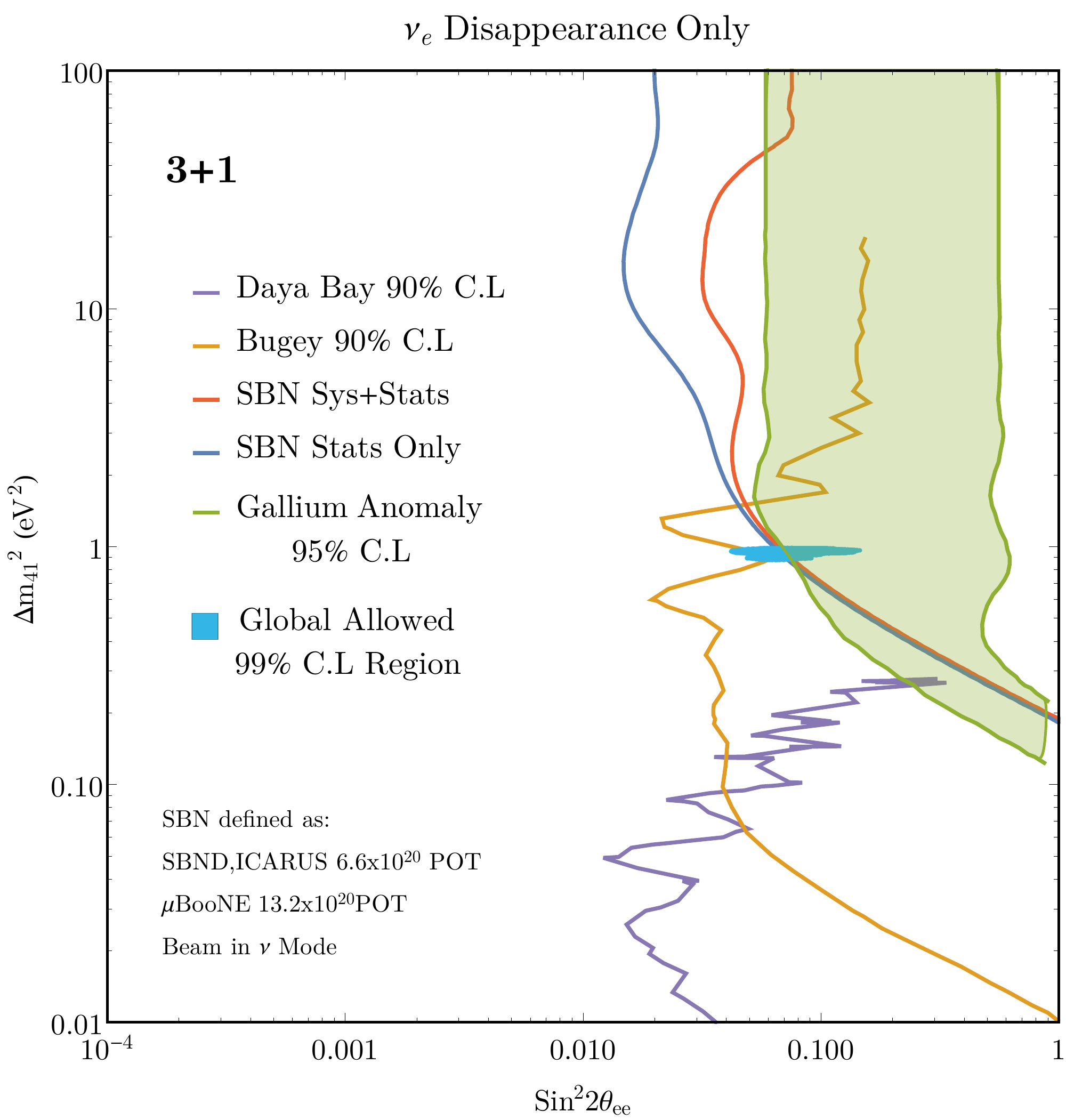}
  \caption{Due to the very large intrinsic $\nu_e$ component of the beam at SBND, one can also perform a $\nu_e$ disappearance only analysis directly probing $\sin^2 2 \theta_{ee}$ at high $\Delta m^2 \geq 0.2 \text{eV}^2$. This is traditionally probed using reactor antineutrinos at a much lower MeV scale energy, and so would provide yet another way of probing the low-energy sterile neutrino anomalies.This is a direct probe of $\sin^2 2 \theta_{ee}$ using a neutrino beam rather than the lower energy (MeV) reactor antineutrinos. } 
  \label{fig:3p1_dise_only}
\end{figure}

Due to the proximity of the SBND experiment to the BNB target, the flux of intrinsic $\nu_e$ at the detector is extremely large. Specifically, SBND expects to record over 35,000~$\nu_e$ CC events for 6.6e20~POT. This allows for an additional oscillation channel to be probed, that of $\nu_e$ disappearance. The SBN $\nu_e$ disappearance-only sensitivity reach is shown in Fig.~\ref{fig:3p1_dise_only} (red curve). We note that this is the first time that SBN's sensitivity to $\nu_e$ disappearance has been explicitly quantified. Although this search is less sensitive to the 1~eV$^2$ region, due to the fact that the $\nu_e$ flux has a relatively high mean energy, at higher $\Delta m^2_{41}$ values it is comparable in $\sin^2 2\theta_{ee}$ reach to that of reactor short-baseline $\bar{\nu}_e$ disappearance bounds. It is also a direct probe of $\sin^2 2 \theta_{ee} $ using a high-energy neutrino beam in complementarity with the MeV-scale antineutrino reactor flux searches. 

Although instructive, none of the above three cases are actually appropriate for an SBN oscillation search if one assumes that the sterile neutrino contains mixing to both the electron and muon sectors. Instead, a proper search for oscillations at SBN should consider the simultaneous effects of both $\nu_e$ disappearance and $\nu_{\mu}$ disappearance and, consequently, $\nu_e$ appearance. We therefore adopt this case, referred to as $\nu_e$ dis/appearance and $\nu_\mu$ disappearance, as the proper SBN sensitivity search method, and we present results corresponding to this case throughout the following sections.

As the primary physics goal of the SBN programme is to \emph{definitively} probe the light sterile neutrino sector that could be responsible for the low-energy anomalies, we use the new metric defined in previous sections to quantify how well SBN can achieve this goal under each of the (3+1), (3+2) and (3+3) scenarios. This metric is referred to as {\it Global $X$\% CL ``Coverage''}, and it refers to the fraction of hypervolume of the $X$\% CL globally-allowed oscillation parameter space that can be ruled out by SBN with a certain confidence level, if SBN observed no oscillations. To estimate global coverage, we first discretize the sterile neutrino parameter space in 100 points in each independent mass-squared difference, mixing element, and CP phase. The mass-squared differences are each discretized over the range of 0.01~eV$^2$ to 100~eV$^2$ (in grid points that are equidistant in logarithmic scale), while the mixing elements $|U_{\alpha i}|$ are discretized in 100 linearly spaced grid points ranging from 0 to 0.5, and the CP-violating phases in 100 points ranging linearly from 0 to $2\pi$. This allows to calculate a hypervolume represented by the number of space points or the ``\emph{size}'' of parameter space that is preferentially allowed by global data at a given confidence interval (in our case, 99\%). We can then express SBN's sensitivity reach as the fractional number of space points or fraction of this hypervolume that SBN can exclude at any given confidence level.

\begin{figure}[h!]
  \centering
  \includegraphics[width=0.45 \textwidth]{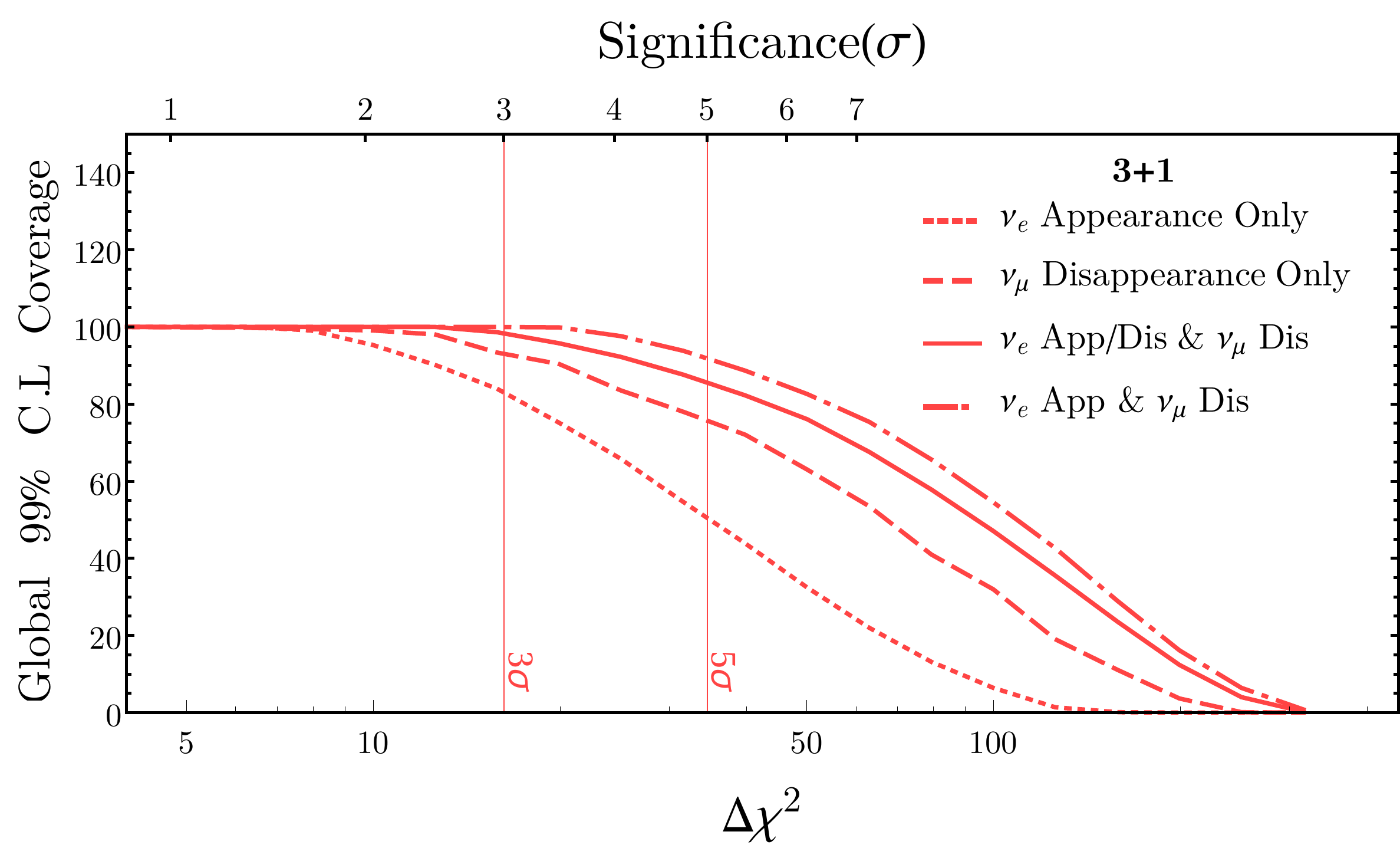}\\
  \includegraphics[width=0.45 \textwidth]{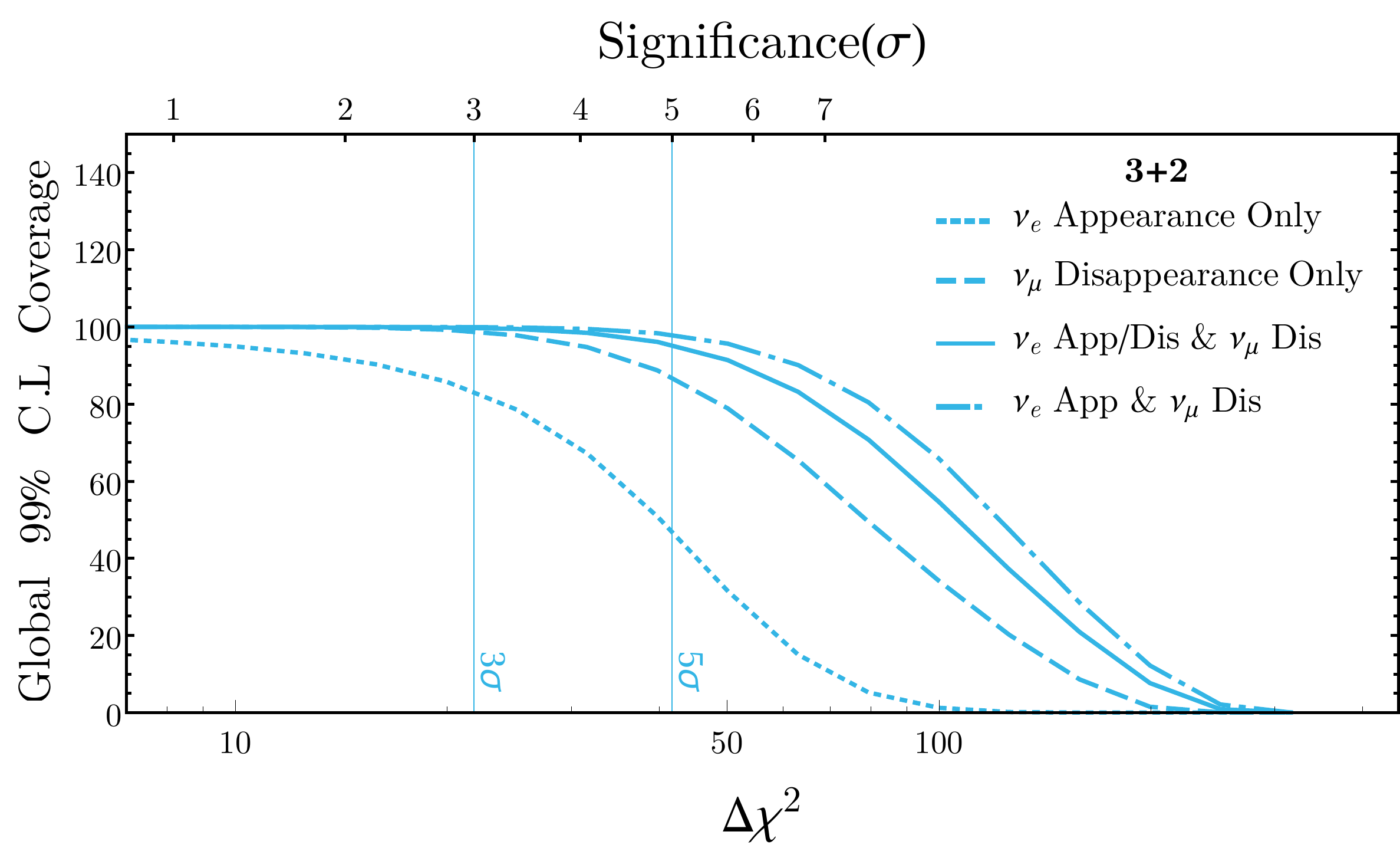}\\
  \includegraphics[width=0.45 \textwidth]{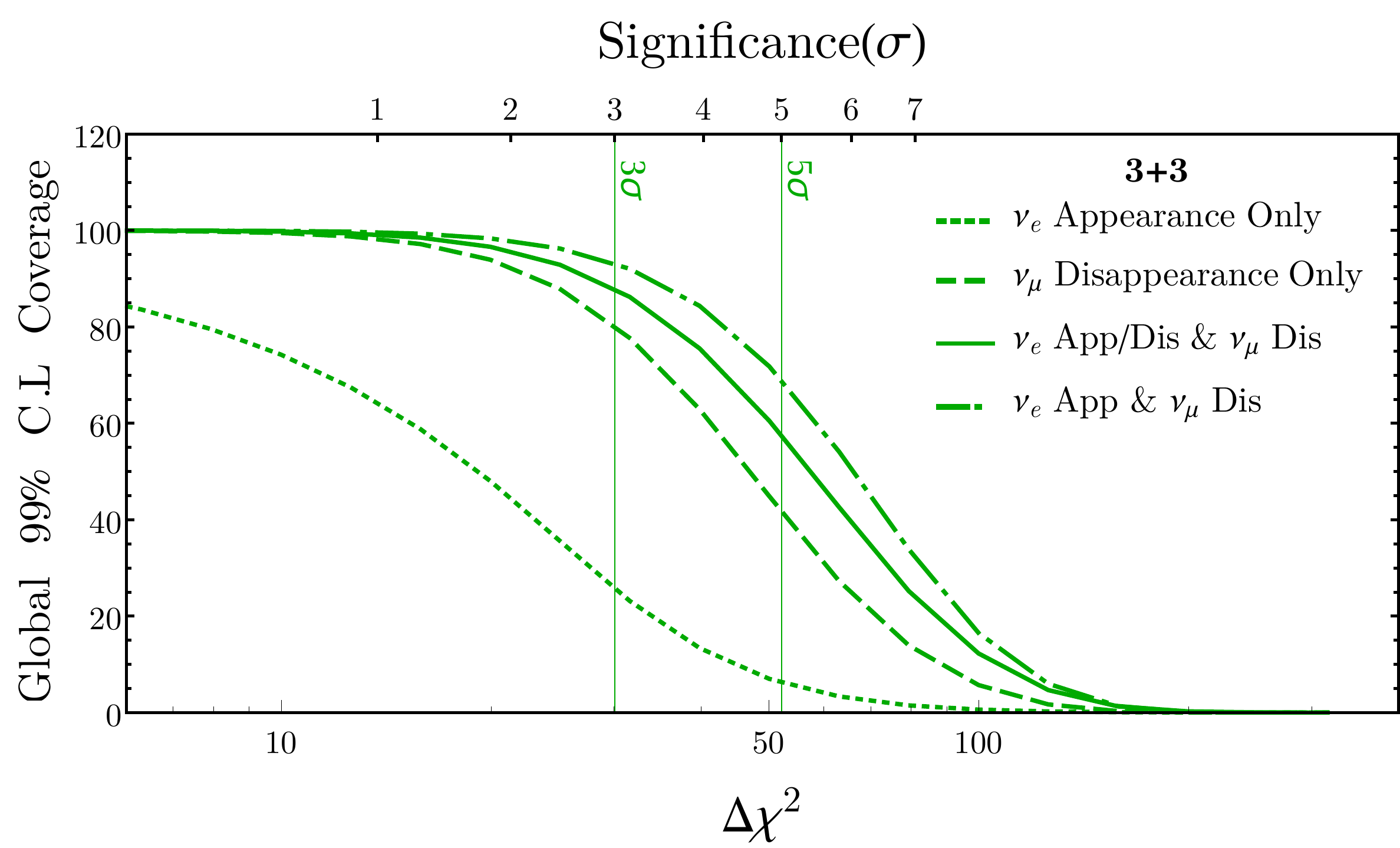}
  \caption{SBN coverage, showing the fraction of 99\% CL allowed global fit region that SBN can exclude at any given $\Delta \chi^2$, for the (3+1) (red, top) (3+2) (blue, middle) and (3+3) (green, bottom) sterile neutrino oscillation scenarios. The dotted curves correspond to $\nu_e$ appearance only searches, The dashed curves correspond to $\nu_\mu$ disappearance only searches, and the solid curves correspond to a combined $\nu_e$ dis/appearance and $\nu_\mu$ disappearance search, which provides the highest sensitivity overall. The percentage covered is shown as a function of $\Delta \chi^2$ on the bottom $x$-axis and as a function of significance on the top $x$-axis, assuming 3, 7 and 12 $dof$ for (3+1), (3+2), and (3+3) fits, respectively.}
  \label{fig:percent_coverage}
\end{figure}

A concrete example of this methodology is shown in Fig.~\ref{fig:percent_coverage}, where we show the percent of the 99\% CL. allowed region that SBN can exclude at a given $\Delta \chi^2$ in a $\nu_e$ appearance only (dotted line), a $\nu_\mu$ disappearance only (dashed line),  as well as a $\nu_e$ dis/appearance and $\nu_\mu$ disappearance (solid line) fit, assuming 6.6e20~POT collected concurrently with all three SBN detectors, after the first MicroBooNE-only run of 6.6e20~POT (with MicroBooNE-only data also included). The results for the (3+1) scenario are shown in the top panel. Shown also are the results for the (3+2) and (3+3) scenarios, in the middle and bottom panels, which will be discussed in their respective sections below.

From the top panel, it is evident that the best performance is possible in the case of a $\nu_e$ dis/appearance and $\nu_\mu$ disappearance search (solid line). In that case, SBN can cover close to 100\% of the 99\% CL globally allowed (3+1) parameter space at 3$\sigma$, and similarly 85\% of the parameter space at 5$\sigma$. In contrast, an appearance-only search can only cover 85\% of the parameter space at 3$\sigma$, and only 50\% of the parameter at 5$\sigma$. We note that in drawing these comparisons we use $\Delta\chi^2$ cuts corresponding to three (3) $dof$ for all three cases ($\nu_e$ appearance, $\nu_\mu$ disappearance, and $\nu_e$ dis/appearance and $\nu_{\mu}$ disappearance).

\begin{figure}[h!]
\begin{center}
  \includegraphics[width=0.45 \textwidth]{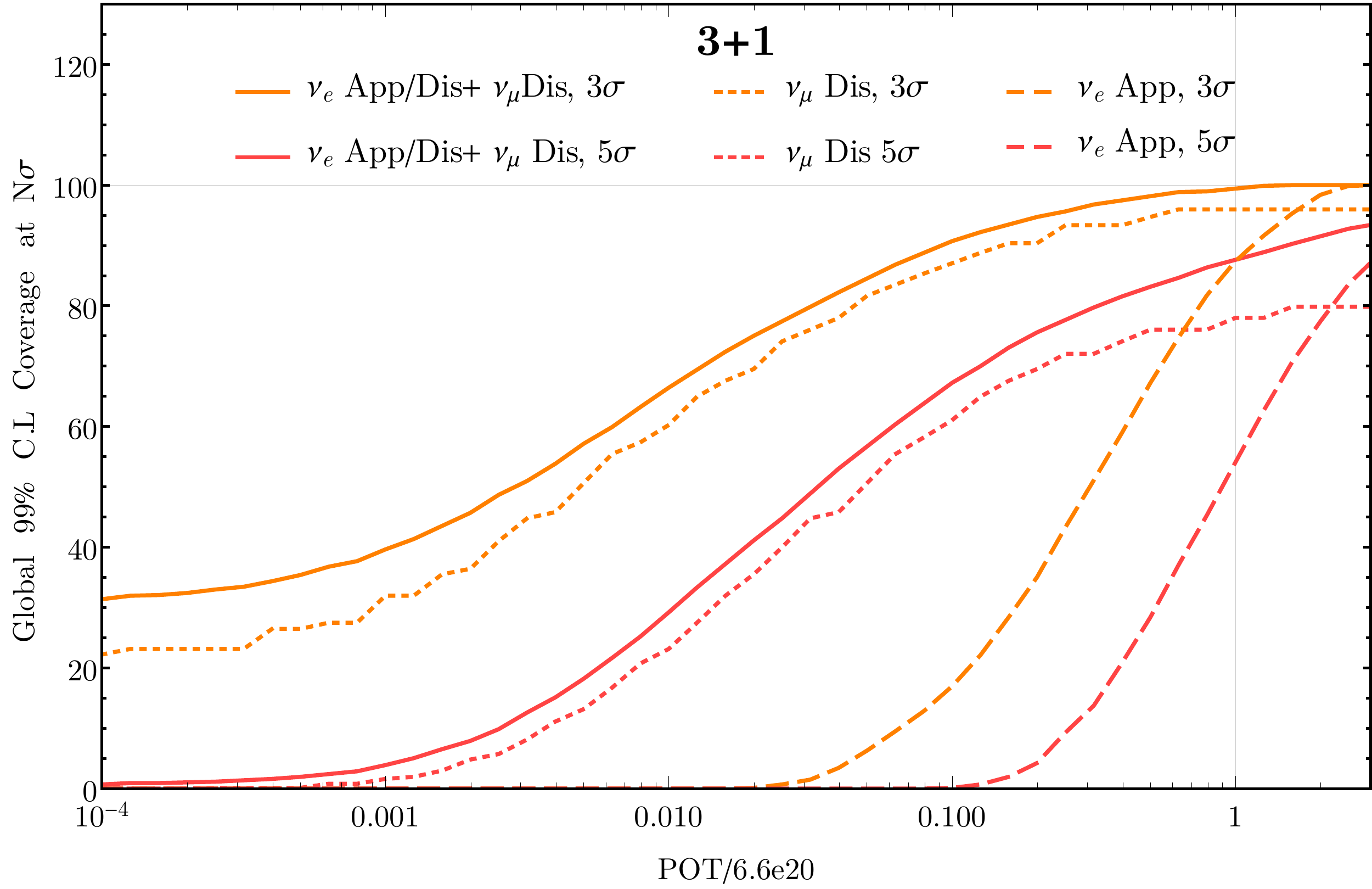}
  \caption{The percentage of 99\% CL globally allowed (3+1) parameter space that SBN can exclude at the 3$\sigma$ (orange) and 5$\sigma$ (red) CL for $\nu_e$ appearance only (dotted), $\nu_\mu$ disappearance only (dashed) and a combined appearance and disappearance fit (solid), as a function of POT.} 
  \label{fig:3p1_pot_coverage}
  \end{center}
\end{figure}

Nevertheless, although a $\nu_e$ dis/appearance and $\nu_\mu$ disappearance search provides a more powerful sensitivity to the (3+1) parameter space, one would like to see a strong exclusion in both the exclusive $\nu_e$ appearance search and the exclusive $\nu_\mu$ disappearance and $\nu_e$ disappearance searches individually in order to conclusively rule out any light sterile neutrino oscillation hypothesis. The POT at which such a statement can be made is explored in Fig.~\ref{fig:3p1_pot_coverage}, which shows the SBN 3$\sigma$ and 5$\sigma$ coverage (in yellow and red, respectively) as a function of POT delivered to the SBN program. As we assume that MicroBooNE has already ran for 6.6e20~POT by the time that the three-detector SBN program commences, the $x$ axis corresponds explicitly to the POT delivered for the three-detector operations, and the plot by construction demonstrates the MicroBooNE-only (6.6e20~POT) coverage at $x=0$. We note that even a MicroBooNE-only combined $\nu_e$ dis/appearance and $\nu_\mu$ disappearance search would yield a 3$\sigma$ coverage of 25\% of the (3+1) globally-allowed parameter space. In general, the total coverage is driven primarily by the $\nu_\mu$ disappearance channel, as evident by the dotted line(s) lying close to the solid line(s).

\begin{figure}[h!]
  \centering
  \includegraphics[width=0.45 \textwidth]{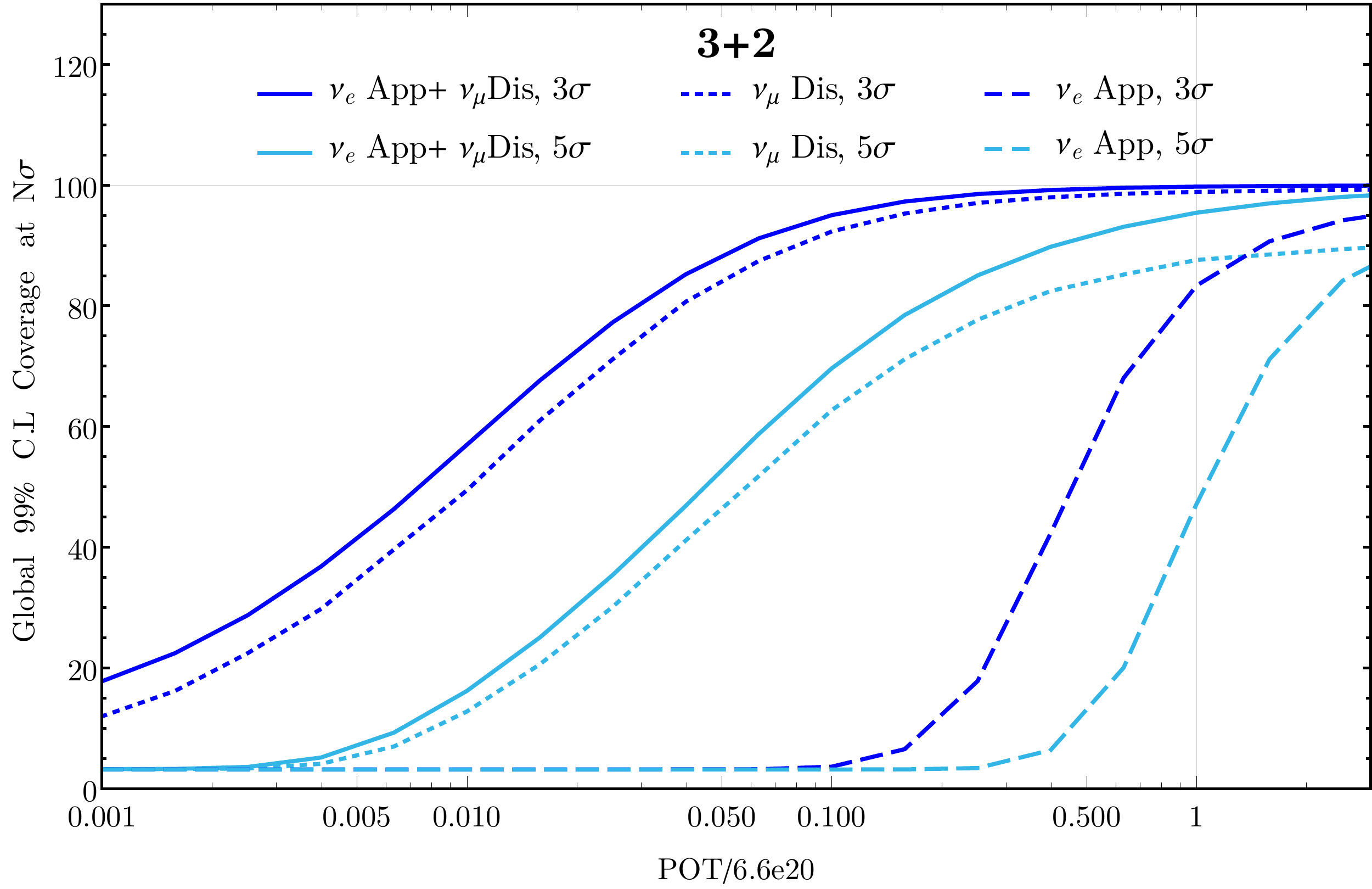}
  \caption{ The same as Fig.~\ref{fig:3p1_pot_coverage} but for the 3+2 light sterile neutrino scenario. The percentage of 99\% CL globally allowed 3+2 parameter space that SBN can exclude at the 5$\sigma$ (light blue) and 3$\sigma$ (dark blue) CL for $\nu_e$ appearance only (dotted), $\nu_\mu$ disappearance only (dashed) and a combined appearance and disappearance fit (solid), as a function of POT. }
  \label{fig:3p2_pot_coverage}
\end{figure}

\subsection{(3+2) Scenario at SBN}
To achieve its goal of definitively addressing sterile neutrino oscillations, SBN will need to have extensive coverage of the (3+2) (and similarly (3+3)) sterile neutrino oscillation parameters as well. In the case of the (3+2) scenario, the additional parameters introduced when one adds another light sterile neutrino happen to enlarge the size of the parameter space that is preferred by the global fits. Nevertheless, as can be seen in the middle panes of Fig.~\ref{fig:percent_coverage}, the percentage of globally allowed (3+2) parameter space (at 99\% CL.) that SBN can cover at any given confidence level is generally comparable to that of the (3+1) scenario. SBN is able to cover 100\% (95\%) of parameter space the 3(5)$\sigma$ level in a combined $\nu_e$ dis/appearance and $\nu_\mu$ disappearance under the (3+2) scenario. In contrast, using $\nu_e$ appearance-only fits, SBN is limited to a maximum of 82(46)\% possible coverage at 3(5)$\sigma$, assuming a nominal exposure of 6.6e20~POT. The SBN 3$\sigma$ and 5$\sigma$ coverage of the (3+2) parameter space as a function of POT can be shown in Fig.~\ref{fig:3p2_pot_coverage}. We note that in drawing these comparisons we use $\Delta\chi^2$ cuts corresponding to seven (7) $dof$ for all three cases.

\begin{figure}[t!]
\begin{center}
  \includegraphics[width=0.45 \textwidth]{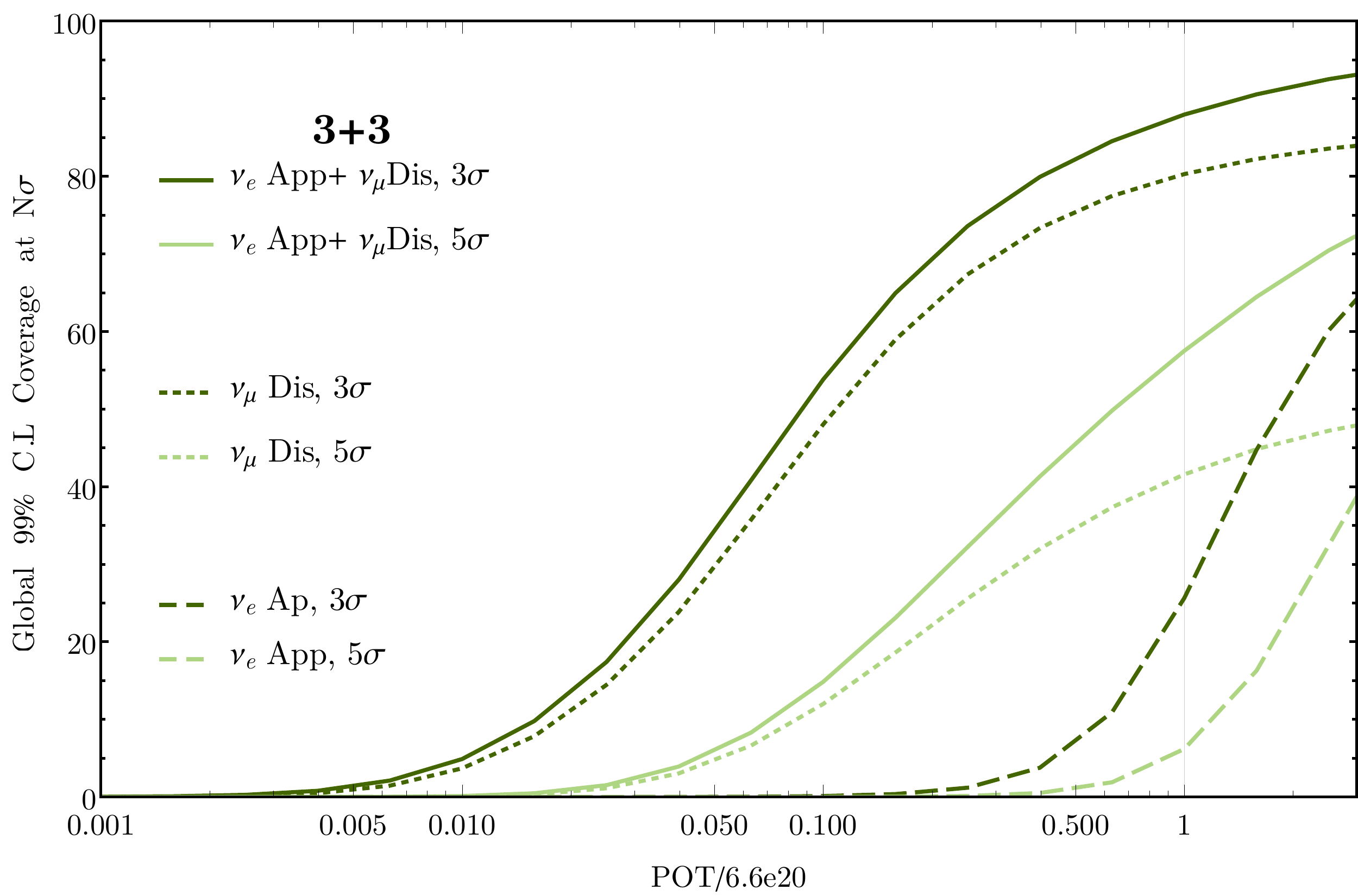}
  \caption{The same as Fig.~\ref{fig:3p1_pot_coverage} but for the 3+3 light sterile neutrino scenario. The percentage of 99\% CL globally allowed 3+3 parameter space that SBN can exclude at the 5$\sigma$ (light green) and 3$\sigma$ (dark green) CL for $\nu_e$ appearance only (dotted), $\nu_\mu$ disappearance only (dashed) and a combined appearance and disappearance fit (solid), as a function of POT. }
  \label{fig:3p3_pot_coverage}
  \end{center}
\end{figure}

\subsection{(3+3) Scenario at SBN}
The (3+3) scenario represents the most challenging scenario for the SBN program to definitively rule out, containing a total of three independent CP-violating phases and twelve independent mass and mixing parameters. As can be seen in Fig.~\ref{fig:percent_coverage}, bottom panel, at its full planned exposure of 6.6e20~POT, the SBN program can cover only 90(57)\% of the globally allowed 99\% CL region at greater than 3(5)$\sigma$, and only with a combined $\nu_e$ dis/appearance and $\nu_\mu$ disappearance search. In a $\nu_e$ appearance-only search, SBN only covers ~25(5)\% of the globally allowed parameter space at 3(5)$\sigma$. The SBN coverage of (3+3) allowed regions as a function of delivered POT is shown in Fig.~\ref{fig:3p3_pot_coverage}. The figure also shows that MicroBooNE alone cannot probe any (3+3) parameter space.

\subsection{$\nu_e$ disappearance effects at SBN}
As this is the first time that SBN's sensitivity to $\nu_e$ disappearance has been demonstrated, we find it interesting to consider explicitly the effect of ignoring $\nu_e$ disappearance effects in the measured $\nu_e$ CC spectra, when performing combined $\nu_e$ appearance and $\nu_\mu$ disappearance fits. We additionally show, in Fig.~\ref{fig:percent_coverage}, the SBN coverage under the (3+1), (3+2), and (3+3) scenarios in a combined $\nu_e$ appearance and $\nu_{\mu}$ disappearance only search (dot-dashed line). By comparing this to the scenario in which the $\nu_e$ background is allowed to oscillate away, it is evident that performing an SBN search for sterile neutrino oscillations without the explicit assumption of negligible disappearance of intrinsic $\nu_e$ backgrounds has a significant effect on SBN's sensitivity, and warrants its consideration along with careful consideration of systematic correlations among exclusive samples measurable at SBN. 

\section{CP-violating phases at SBN}\label{sec:cp}
The addition of CP-violating phases in the (3+2) and (3+3) sterile neutrino scenarios introduces the potential of new oscillation probability asymmetries at SBN that would be observable in comparisons of neutrino and antineutrino oscillations.  Although there is currently no planned antineutrino running for SBN, when considering the possibility non-zero CP-violating phases associated with sterile neutrinos, it is natural to ask whether SBN's sensitivity coverage improves with the inclusion of a combination of neutrino and antineutrino running. In particular, one may consider whether SBN's ability to rule out the short-baseline anomalies improves with the addition of antineutrino running. Another consideration is whether additional antineutrino running would allow for more precise measurements of new neutrino mass splittings and mixings and in particular any CP-violating phases associated with $N$ additional states, should a potential sterile neutrino signal be confirmed with SBN neutrino running. 

\subsection{Antineutrino coverage in the absence of a signal}\label{sec:antineutrino}
To investigate the impact of antineutrino running at SBN, we expand the fit as described in Sec.~\ref{sec:simulation} to include observable $\nu_e$ CC and $\nu_{\mu}$ CC spectra at the three SBN detectors in antineutrino running mode, as well as in neutrino mode. The same background definitions are considered as in neutrino mode, and the backgrounds are re-evaluated assuming no right- or wrong-sign discrimination within each event sample, as described in Sec.~\ref{sec:spectra}. 

\begin{figure}[h!]
  \centering
  \includegraphics[width=0.45 \textwidth]{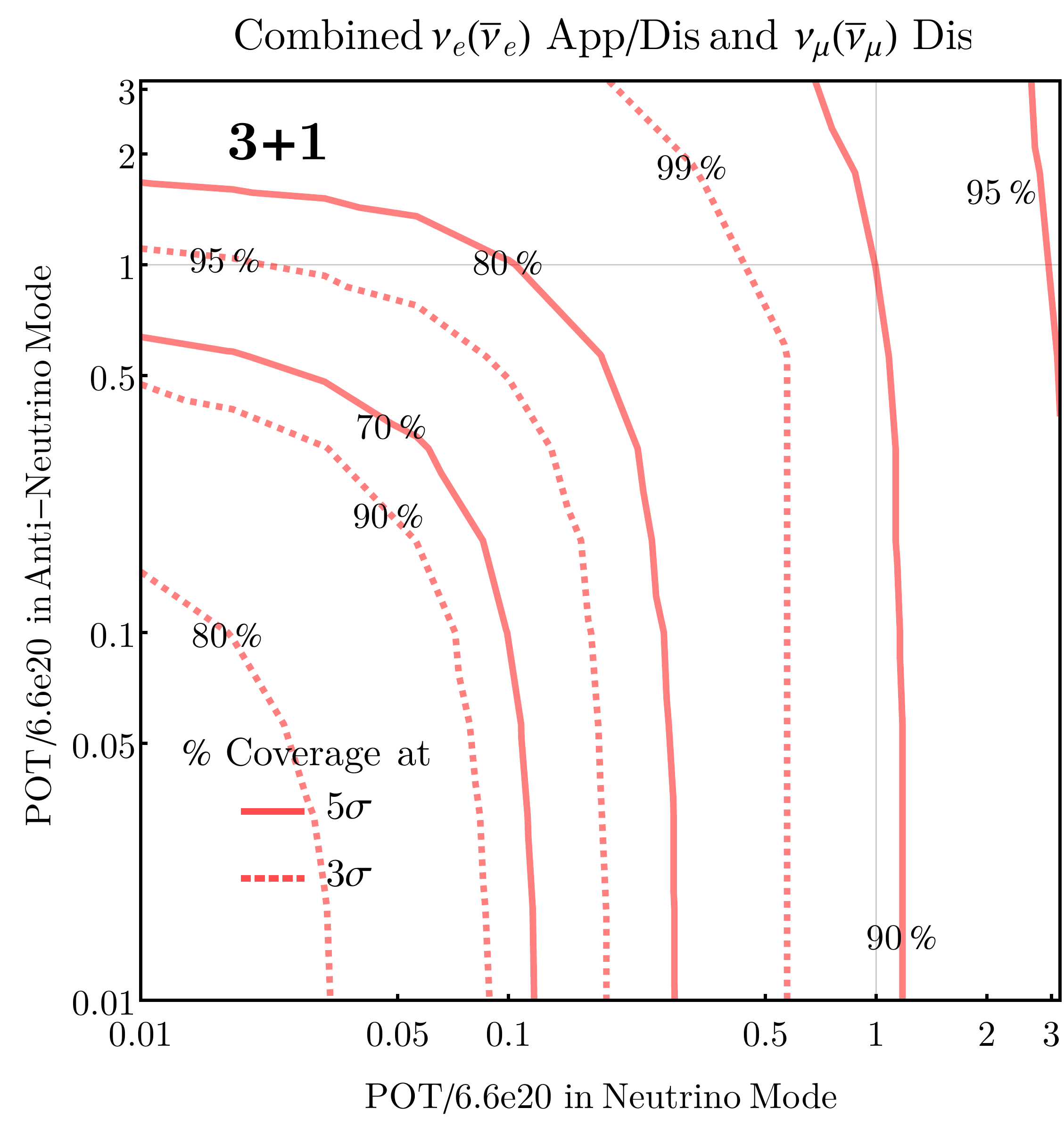} \\
  \begin{raggedleft}
  \caption{The amount of POT required in neutrino and antineutrino running modes for SBN to cover a given percentage of the 99\% CL globally allowed regions at 3$\sigma$ (dashed curves) and 5$\sigma$ (solid curves) in the (3+1) light sterile neutrino scenario. This corresponds to a combined $\nu_e$ dis/appearance and $\nu_\mu$ disappearance search. Note that, as MicroBooNE will have already collected 6.6e20~POT in neutrino mode before SBN begins its run, the $x$-axis refers to additional POT beyond this 6.6E20~POT collected for MicroBooNE-only neutrino mode running.}
  \label{fig:3p1_nuVnubar}
  \end{raggedleft}
\end{figure}

First, coverage is evaluated for a variety of additional beam exposures (beyond the first 6.6e20~POT in neutrino running mode). Figure~\ref{fig:3p1_nuVnubar} shows the exposure in POT for additional neutrino and additional antineutrino running (and combinations) that the SBN program requires, in order to probe the 99\% CL globally allowed regions at 3$\sigma$ (solid curves) and 5$\sigma$ (dashed curves) for the (3+1) scenario at a percentage coverage as indicated explicitly on each curve. We focus on the strongest exclusion case, as motivated in Sec.~\ref{sec:results3p1}, corresponding to a combined $\nu_e$ dis/appearance and $\nu_\mu$ disappearance fit. We highlight that it is far more efficient to cover a given fraction of parameter space with additional neutrino-only running, rather than antineutrino-only or any combination of additional neutrino plus antineutrino running. This is evident from these figures as no point on any curve deviates from the origin by a distance smaller than the curve's $x$-coordinate for $y=0$. This is expected for the (3+1) scenario, as neutrino and antineutrino oscillation probabilities under the two-neutrino oscillation approximation we've employed are identical by construction. Therefore, antineutrino running offers no additional information, and it is generally less efficient due to the lower flux and cross-section, and, hence, event statistics.

\begin{figure}[h!]
  \centering
  \includegraphics[width=0.45 \textwidth]{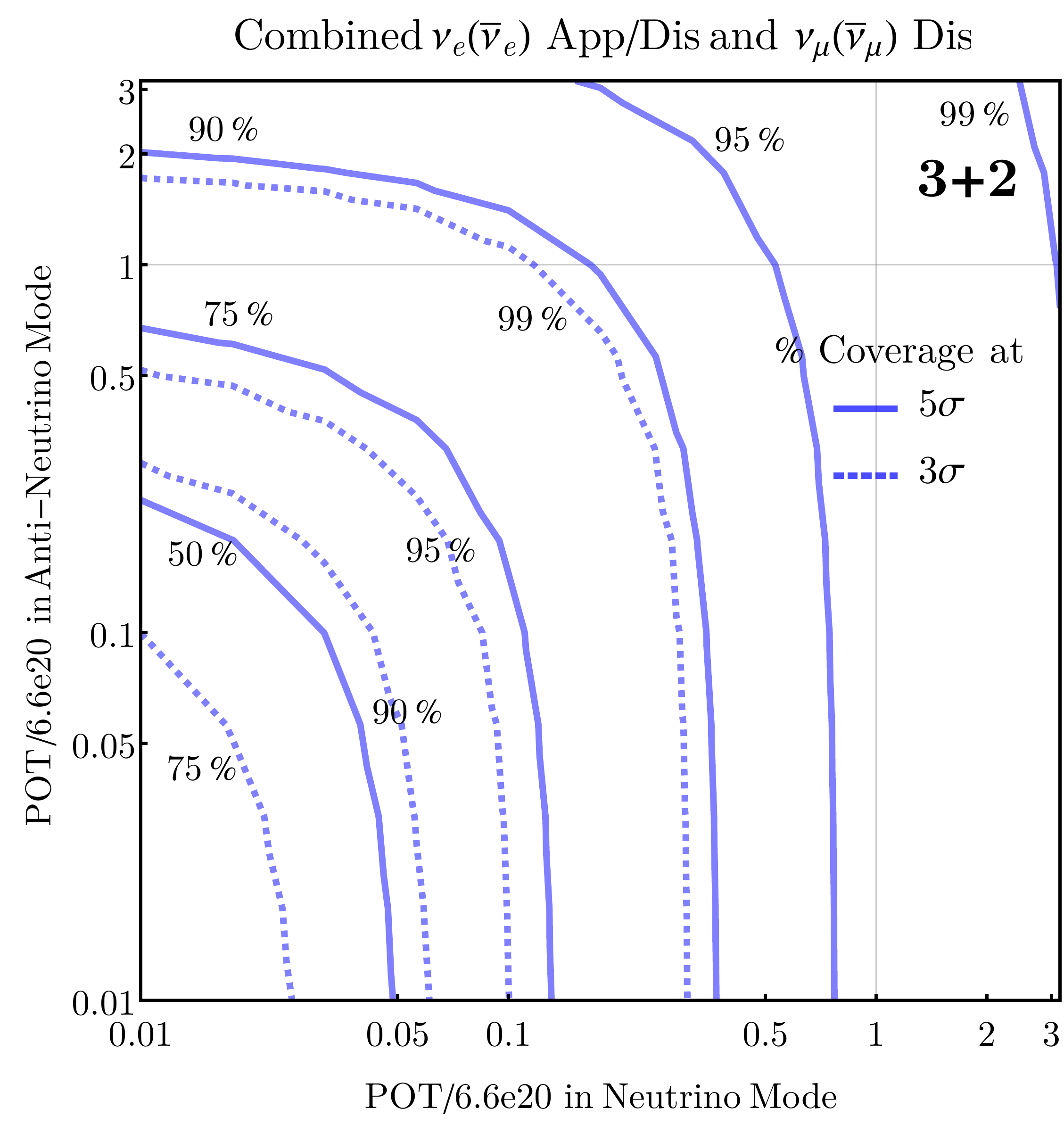} \\
  \caption{The same as Fig.~\ref{fig:3p1_nuVnubar} but for the (3+2) light sterile neutrino scenario}  \label{fig:3p2_nuVnubar}
\end{figure}

\begin{figure}[h!]
  \centering
  \includegraphics[width=0.45 \textwidth]{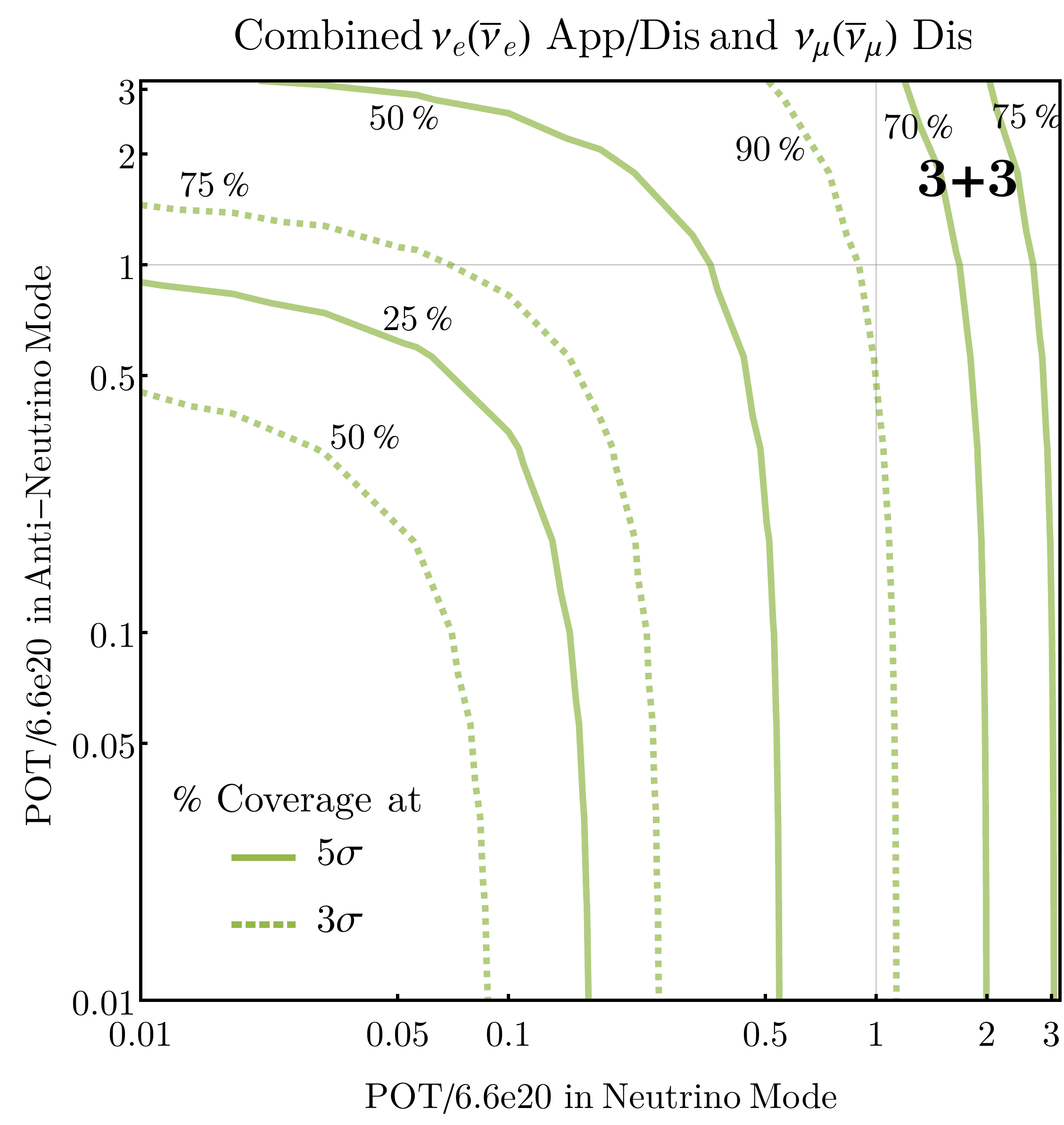} \\
  \caption{The same as Fig.~\ref{fig:3p1_nuVnubar} but for the (3+3) light sterile neutrino scenario }  \label{fig:3p3_nuVnubar}
\end{figure}

Figures~\ref{fig:3p2_nuVnubar} and \ref{fig:3p3_nuVnubar} show the same information for the (3+2) and (3+3) scenarios, respectively. Interestingly, just as in the (3+1) case, we observe again that it is far more efficient to cover any given fraction of parameter space with additional neutrino-only rather than antineutrino-only or any combination of neutrino plus antineutrino running. At first this may seem counter-intuitive, as it may be expected that antineutrino running would provide visibly more coverage due to enhanced sensitivity to CP-violating phases in these scenarios. However, the increased statistics per POT that are available in neutrino mode running are far more efficient in constraining all other mixing parameters and masses allowed in these oscillation hypotheses. Since these plots quantify overall coverage of the $n$-dimensional phase-space in each scenario, it is quite reasonable (and arguably expected) that antineutrino running proves less effective in terms of this metric. 

In the absence of a possible signal, additional POT in antineutrino mode (as opposed to neutrino mode) does \emph{not} help to rule out the null hypothesis faster, for any scenario. It can be argued that SBN's sensitivity to CP violation through comparisons of neutrino and antineutrino running spectra suffers from the significant\footnote{Approximately 30\% of events in antineutrino running mode are expected to be due to neutrino interactions.} wrong-sign neutrino contribution inherent in the BNB beam when running in antineutrino mode. Further studies into methods of differentiating between neutrino and antineutrino events in a LArTPC, such as exploiting $\mu^-$ absorption rates on argon or the difference in $Q^2$ distributions of $\nu$ and $\overline{\nu}$ interactions, would be especially useful in quantifying the impact on SBN's sensitivity to CP violation, and also of interest to LArTPC development in general. It would also be worthwhile for SBN to consider whether any BNB optimization is possible and could be implemented to minimize the wrong-sign flux.

On the other hand, if SBN observes a sterile neutrino-like signal in neutrino running mode, the focus would quickly turn to the subsequent measurement of the new parameters. Here, the impact of SBN antineutrino running may become important, providing access to a potentially distinctly different observable oscillation probability than the neutrino run would allow. However, the challenge is that the CP-violating phase effects become degenerate with those of the remaining oscillation parameters, in particular with insufficient detector energy resolution. In what follows, we explore this possibility, but we focus solely on the (3+2) scenario with a single CP-violating phase $\phi_{54}$, for simplicity; however, these metrics could be applied to the (3+3) scenario with minimal expansion.

\subsection{Sensitivity to $\phi_{54}$ }\label{sec:cpsens}
The sensitivity of SBN to the CP-violating phase $\phi_{54}$ is studied under the hypothesis that SBN observes a signal consistent with two light sterile neutrinos. To analyse this sensitivity  we inject potential signals, for a given set of oscillation parameters, into the fit. These injected parameters are labelled as ``true'' parameters, and the spectra produced when one assumes these parameters take the place of the ``null'' spectra in the $\chi^2$ calculation and covariance matrix construction as described in Sec.~\ref{sec:simulation}. 
This $\chi^2$ quantifies SBN's ability to confirm a certain set of oscillation parameters given a hypothetical signal. Sensitivity to $\phi_{54}$ is due solely to the $\nu_e$-appearance channel in which it uniquely appears. Due to this as well as the large number of degrees of freedoms in the (3+2) sterile neutrino scenario, we make here the simplifying assumption that $|U_{e4}|^2|U_{\mu4}|^2 = |U_{e5}|^2|U_{\mu5}|^2$ and analyse under the assumption of $\nu_e$ appearance only, so as to better understand and convey the behaviour in 2D of the main parameter of interest, $\phi_{54}$. Although allowing all parameters to vary uniquely does indeed change the quantitative results, the qualitative phenomenology remains consistent.

\begin{figure}[t!]
  \centering
  \includegraphics[width=0.45 \textwidth]{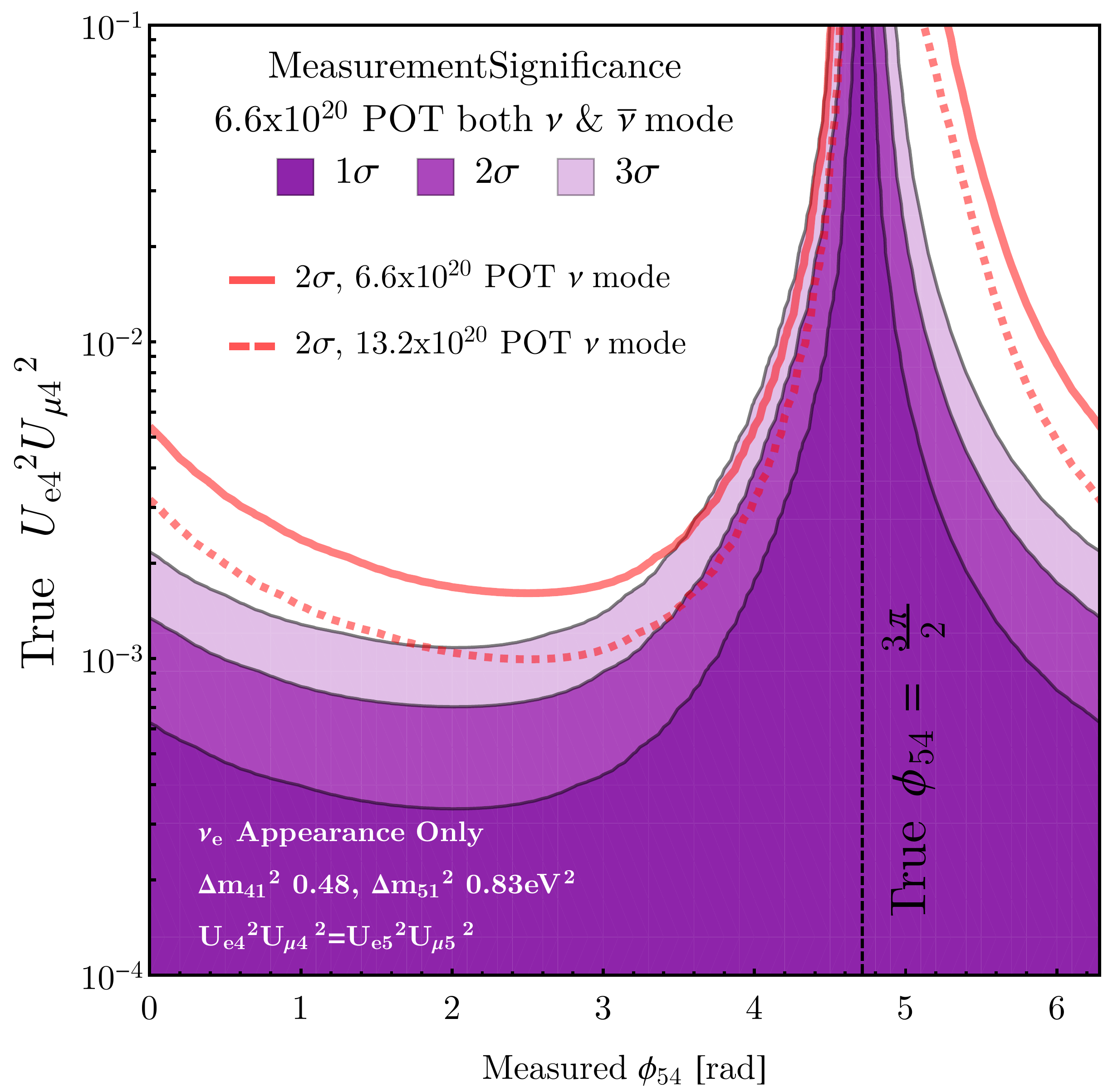}
  \caption{
	  Sensitivity of SBN to a (3+2) scenario sterile neutrino signal, as a function of true mixing $U_{e4}^2 U_{\mu 4}^2$ for $\phi_{54}^\text{true} = 3\pi/2$. We show the regions of reconstructed $\phi_{54}$ that is in agreement at 1,2 and 3$\sigma$ significance in purple shaded regions for a combined 6.6e20~POT neutrino running mode and 6.6e20~POT antineutrino running mode. In dashed red we also show the equivalent 2$\sigma$ contour for 13.2e20~POT neutrino running mode only. The mass splittings correspond to the global (3+2) best fit point. As the true mixings are fixed in each test case, the contours are drawn at $\Delta \chi^2$'s of  1,4 and 9, corresponding to the 1 remaining $dof$, $\phi_{54}$, after profiling over all other parameters.
	} \label{fig:phi54_sens_4.7_bf}
\end{figure}

In Fig.~\ref{fig:phi54_sens_4.7_bf}, we show a sample scenario in which we inject a true $\phi_{54}$ of $3 \pi/2$, for values of mass splittings from our simulated grid, chosen to be closest to the global best fit, $\Delta m_{41}^2 = 0.48$~eV$^2$ and $\Delta m_{51}^2 = 0.83$~eV$^2$. We then vary the strength of the active neutrino-sterile neutrino mixings, $|U_{e4}|^2|U_{\mu 4}|^2$ and show the range of possible reconstructed $\phi_{54}$ values which fit the injected signal within a given confidence level, all the while marginalizing over remaining mixing elements.

For a mass splitting of $\sim 1$~eV$^2$, explaining the LSND anomaly requires mixings of order $|U_{e4}|^2|U_{\mu 4}|^2 \approx 10^{-4} - 2\times 10^{-3}$. We note that $\phi_{54}$ resolution in this region varies from no-sensitivity to $\pm$ 40$^\circ$ at the $1\sigma$ level. Under the standard exposure of 6.6e20~POT in neutrino mode alone (red solid line) one can see there is no sensitivity for even the largest values of mixing parameters consistent with the (3+2) global data, $|U_{e4}|^2 |U_{\mu 4}|^2 \approx 2\times 10^{-3}$. As such, we concentrate on whether of not it is advantageous to run further in neutrino mode (red dashed line) or a combination of neutrino and antineutrino running mode (purple shaded regions). As can be seen, for unrealistically large mixing, SBN can strongly pick out the true $\phi_{54}$, but, as the mixing drops, the resolution on $\phi_{54}$ reduces until one reaches $|U_{e4}|^2|U_{\mu 4}|^2 \approx 4\times10^{-4}$, by which all values of $\phi_{54}$ are indistinguishable. We also show the 2$\sigma$ contour for the case in which we run entirely in neutrino-mode for an additional 6.6e20~POT (red dashed lines) and note that for the majority of the parameter space, is worse than a combined neutrino and antineutrino exposure.

\begin{figure}[t!]
  \centering
  \includegraphics[width=0.45 \textwidth]{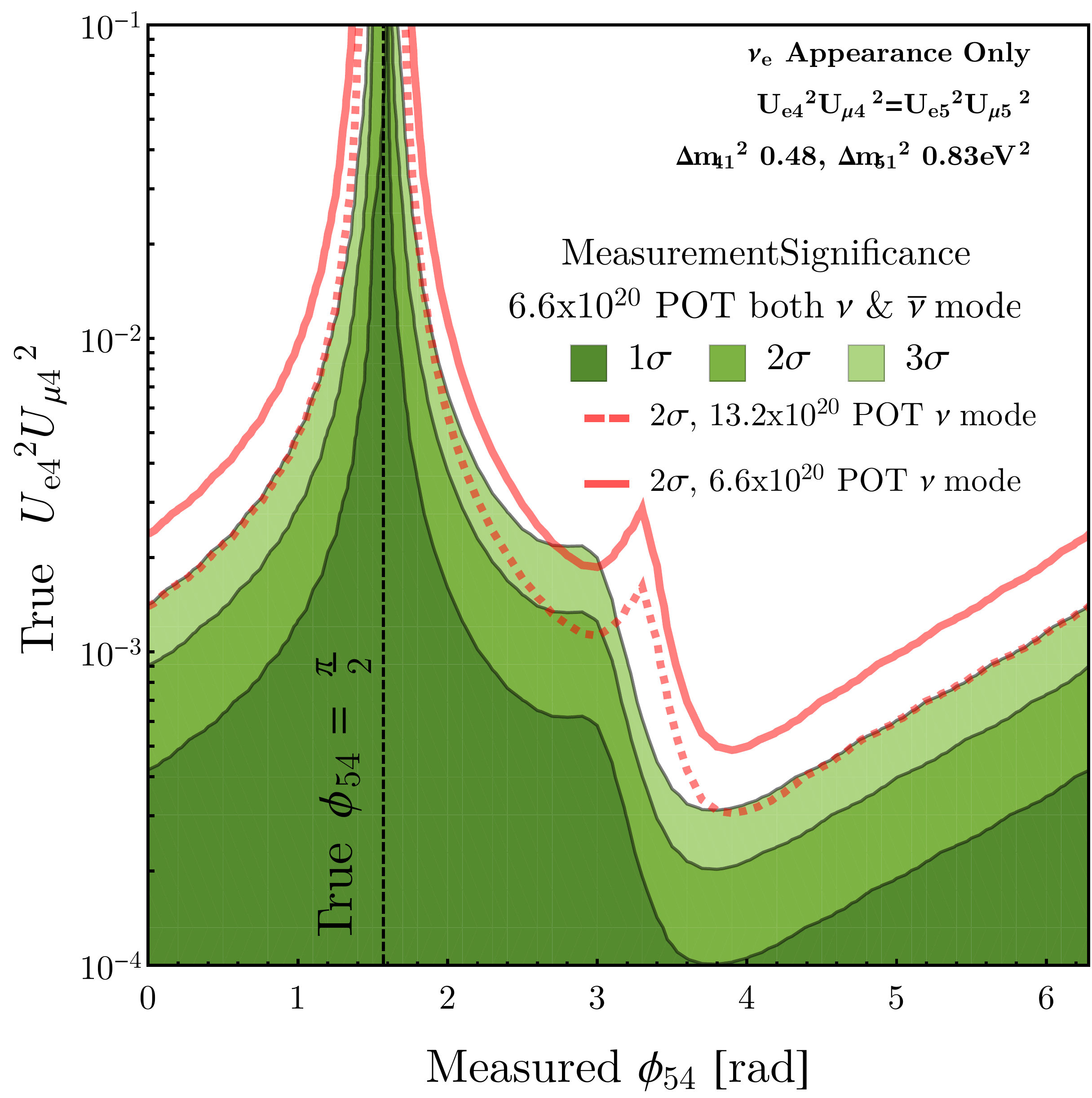}
	\caption{Same as Fig.~\ref{fig:phi54_sens_4.7_bf} but for injected $\phi_{54}^\text{true} = \pi/2$ and mass splittings corresponding to the largest mixing allowed by current global (3+2) best fit point. See text for more details.
	} \label{fig:phi54_sens_1.5_max3}
\end{figure}

The exact sensitivity of $\phi_{54}$ depends not only on the magnitude of mixings, but also on the assumed mass splittings. In Fig.~\ref{fig:phi54_sens_1.5_max3}, we repeat the same analysis for $\phi_{54}=\pi/2$, $\Delta m_{41}^2 = 0.16$ eV$^2$ and $\Delta m_{51}^2 = 1.0$~eV$^2$. This point corresponds to the point with largest mixings allowed in our (3+2) global fit at the 99\% CL. The green shaded region assumes 6.6e20~POT in both neutrino and antineutrino running and shows sensitivity to $\phi_{54}$ for values of values of $U_{e4}^2U_{\mu 4}^2$ as low as $10^{-4}$. Again we see that running in 50:50 neutrino and antineutrino running mode, over pure neutrino mode (red lines), allows one to measure the true value of $\phi_{54}$ with much higher resolution.

\subsection{\label{sec:cpviol}Prospects for CP violation discovery}
A related measurement to that of determining the value of $\phi_{54}$ given an observed signal, is the significance with which SBN could potentially rule out the CP conserving values of $\phi_{54}=0$ or $\pi$. Establishing CP violation in the sterile neutrino sector would be a crucially important discovery in itself, as well as of great relevance to future experiments looking to measure the standard three-neutrino phase $\delta_{CP}$ \cite{Gandhi:2015xza}.
To estimate SBN's reach with respect to this question, for a given injected signal with $\phi_{54}=\phi_\text{true}$ and fixed values of $\Delta m^2_{i1}$ and $|U_{\alpha i}|$, we form the metric
\[
	\chi^2_\text{CP}(\phi_\text{true}) \equiv \text{Min}[\chi^2(\phi_{54}=0 | \phi_\text{true} ), \chi^2(\phi_{54}=\pi | \phi_\text{true} )]~.
	\]
In each $\chi^2(\phi_{54}=x|\phi_{true})$, all active-sterile neutrino mixing elements are varied in order to find the set which minimizes the $\chi^2$ under consideration, to account for possible degeneracies in the observed spectra. To get as realistic a measurement as possible we relax the simplifying constraint that $|U_{e 4}|^2|U_{\mu 4}|^2 =|U_{e 5}|^2|U_{\mu 5}|^2$ and allow all parameters to vary, fitting to a combined $\nu_e$ appearance and $\nu_\mu$ and $\nu_e$ disappearance.

\begin{figure}[h!]
  \centering
  \includegraphics[width=0.45 \textwidth]{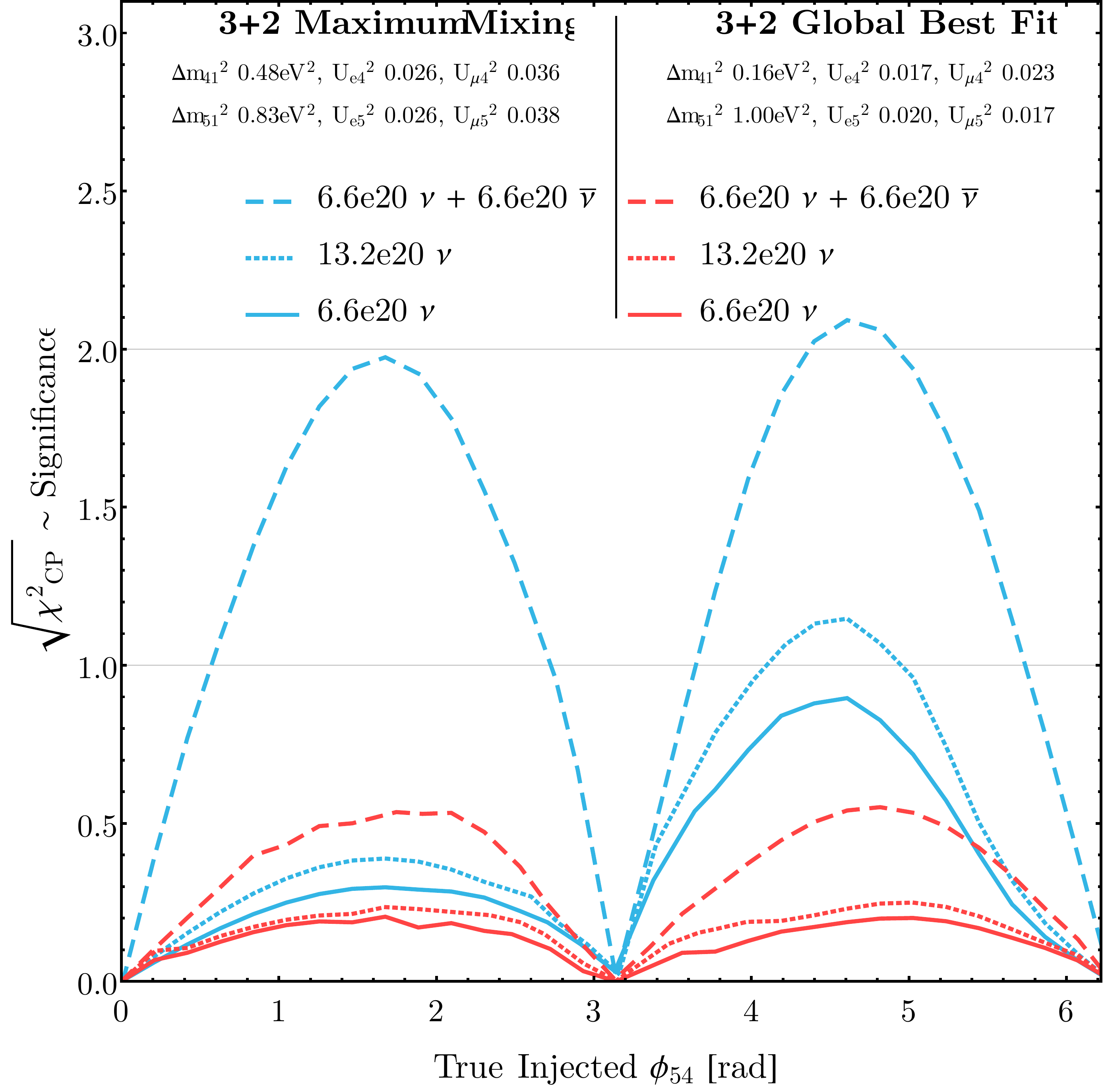}
	\caption{Significance at which SBN can observe CP violation in the (3+2) sterile neutrino scenario, as a function of true $\phi_{54}$, for two injected signals corresponding to the global (3+2) best fit point (red lines) as well as the parameter point with largest total mixings (blue lines), for a variety of POT in neutrino and antineutrino running modes. Unlike previous plots we make no assumption on mixing and fit to $\nu_e$ appearance and both $\nu_e$ and $\nu_\mu$ disappearance simultaneously, profiling over $U_{e4},U_{\mu 4}, U_{e5}$ and $U_{\mu 5 }$. As all remaining parameters are profiled over, and only 1 $dof$ remains, the $\sqrt{\chi^2}$ will approximate the significance of the measurement.
	} \label{fig:CPviolation}
\end{figure}

In Fig.~\ref{fig:CPviolation}, we show the results of this test for the same two possible injected signals as in Figs.~16 and 17---the global (3+2) best fit point (red lines) and the ``maximum allowed mixing'' point (blue lines). For smaller values of mixings, corresponding to the best fit point, little or no spectral shifts can be measured due to varying $\phi_{54}$, and as such even for maximally violating CP angle values, $\phi_{54}$ can always be mis-reconstructed to one of the CP conserving value, with shifts in $|U_{\alpha i}|$ to compensate for the rate. The results for the nominal SBN run plan of 6.6e20~POT in neutrino mode is shown by the solid line and shows no sensitivity to CP violation; similarly, if we assume an additional 6.6e20~POT in neutrino mode, the situation does not change (dotted line) significantly. Although the inclusion of 6.6e20~POT in antineutrino mode (dashed line) does double the potential sensitivity, this remains a sub-$1\sigma$ effect and thus it is clear that within reasonable exposure SBN is completely insensitive to CP violation if nature does choose sterile neutrinos at this mass splitting.\\

As the strength of mixing increases, individual variations in the energy spectrum due to $\phi_{54}$ driven oscillations becomes harder for degeneracies in mixing to explain, and the significance at which certain CP-violating phases are in disagreement with $\phi_{54}=0$ or $\pi$ increases. This is evident when we look at the CP violation curves assuming the ``maximum allowed mixing'' sterile neutrino parameters. If we again assume a standard exposure of 6.6e20~POT in neutrino mode (solid blue line) it is evident that SBN has no sensitivity to CP violation, with significance's of less than $1\sigma$ even with maximum CP violation. Doubling the POT in neutrino mode (dotted blue line) gives an effectively negligible increase, but it is here that the benefits of additional antineutrino running is most evident. An additional 6.6e20~POT in antineutrino mode allows for ~$2\sigma$ significance at maximal mixing, with $>1\sigma$ significance over 68\% of $\phi_{54}$ parameter space. Although certainly not enough to claim discovery, SBN could provide the first hints of CP violation in the sterile neutrino sector in this specific scenario.

It is worth clarifying that even if nature is kind enough to choose a maximally CP-violating phase, $\phi_{54} = \pi/2$ or $3\pi/2$, thus enabling SBN to potentially observe CP violation at the ~2$\sigma$ significance level, it would still require large mixings that are already somewhat in tension with global data $|U_{\mu 5}|^2 \approx 0.0038$, and only certain sterile neutrino mass splittings. For non-maximally violating CP phases the significance at which SBN can make statements diminishes rapidly, and for the majority of the parameter space motivated by the short-baseline anomalies, the potential for SBN to measure a CP-violating phase to the accuracy necessary to rule out CP conservation is very low and insignificant. Conversely, for values of active-sterile neutrino mixings and $\Delta m^2$ splittings outside of those considered here, namely ones which help less to explain the short-baseline anomalies but could be interesting models nonetheless, the sensitivity to CP violation could be significantly greater than those presented here.

\section{Summary and Conclusions}\label{sec:conclusion}

We have considered, for the very first time, SBN's sensitivity to extended light sterile neutrino oscillation scenarios. We find that, in the case of a (3+1) oscillation scenario, SBN is capable of definitively exploring (i.e.~with 5$\sigma$ coverage) 85\% of the 99\% CL parameter space region which is allowed by global short-baseline oscillation data (for 3~$dof$). This is possible after a three year neutrino mode run with all three SBN detectors running concurrently to collect data corresponding to 6.6e20~POT, and with a combined $\nu_e$ dis/appearance and $\nu_\mu$ disappearance search. Furthermore, by performing such a combined search, MicroBooNE alone, during its first three years of running prior to the SBN program commencing, will be able to test 25\% of the globally allowed (3+1) oscillation parameter space at 3$\sigma$.

In the case of a (3+2) scenario, in its planned three-year neutrino run, SBN can definitively explore 95\% of the 99\% CL allowed parameter space (7~$dof$). In this scenario, a single CP-violating phase, $\phi_{54}$, enters in the $\nu_{\mu}\rightarrow\nu_e$ appearance probability and leads to differences in neutrino and antineutrino appearance probabilities. Dedicated BNB antineutrino mode running for three years (6.6e20~POT), beyond the currently planned SBN neutrino mode running, does not significantly expand SBN's 5$\sigma$ sensitivity for ruling out this oscillation scenario. Nevertheless, by performing a multi-baseline and multi-channel oscillation search with sign-selected neutrino and antineutrino beams, the SBN experiment will be able to, within six years of operation, overconstrain a significant fraction of parameter space which is currently allowed by global fits to sterile neutrino oscillation. 

Furthermore, in the case where a potential signal consistent with multiple light sterile-neutrinos is confirmed, dedicated antineutrino running at SBN proves to be of substantial value in increasing the significance of an observation of any CP violation. For the (3+2) sterile neutrino scenario, an additional 6.6e20~POT in antineutrino running mode could allow SBN to provide the first ~$2\sigma$ hints of CP violation in the (extended) lepton sector, provided nature chooses maximal CP-violating phases $\phi_{54} = \pi/2$ or $3\pi/4$, and oscillation parameters consistent with global data at the 99\% CL: $\Delta m_{41}^2 = 0.16$~eV$^2$, $\Delta m_{51}^2 = 1.0$~eV$^2$, $|U_{e4}|^2 = |U_{e5}|^2 =0.026$, $|U_{\mu 4}|^2=0.036$ and $|U_{\mu 5}|^2 = 0.0038$. For SBN to be able to observe CP violation at a greater significance than this would require active-sterile mixing already in significant tension with global data. It is possible that a higher significance could be achieved if the SBN detectors are capable of differentiating between neutrino and antineutrino interactions, either on an event-by-event basis, or through additional statistical treatment of the event samples. Such possibility would be worth exploring through dedicated studies, or through potential beam design upgrades.

In the case of a (3+3) scenario, in its planned three-year neutrino run, SBN can definitively explore 55\% of the currently allowed parameter space. We further note that in all scenarios, (3+1), (3+2), and (3+3), utilizing a simultaneous search for  oscillations in multiple channels ($\nu_e$ appearance, $\nu_e$ disappearance, and $\nu_{\mu}$ disappearance) has a signficant effect on the sensitivity reach. In particular, combining $\nu_e$ and $\nu_\mu$ channels is generally more powerful than exclusive channel searches, except when $\nu_e$ disappearance effects are included in the fit. The latter tend to slightly degrade the sensitivity, due to added degeneracies of the effectively opposite $\nu_e$ appearance and $\nu_e$ disappearance effects. As such, it would be prudent for SBN to carry out a multi-channel search that accounts for all three effects simultaneously.

Finally, we must point out a caveat in these studies, in that the experimental data sets used to constrain the (3+$N$) oscillation parameter suffer from large apparent incompatibility within the parameter space they seem to prefer. Still, we consider it a more conservative approach to consider the globally-allowed rather than the anomaly-allowed region in exploring SBN's discovery reach in terms of fractional coverage of allowed parameter space. 

\begin{acknowledgements}
The authors would like to thank M. Shaevitz, L. Camilleri, B. Louis and S. Pascoli for valuable discussions. This work has been supported in part by a 2014 Institute for Particle Physics Phenomenology (IPPP) Associateship. We also acknowledge support by the Science and Technology Facilities Council (UK). MRL acknowledges partial support from the European Union FP7 ITN INVISIBLES (Marie Curie Actions, PITN-GA-2011-289442).
\end{acknowledgements}

\end{document}